\documentclass[amsmath,amssymb,twocolumn,aps,prb]{revtex4-2}

\usepackage{
	bm,
	dotlessi,
	graphicx,
	hyperref
}

\graphicspath{{figures/}}

\allowdisplaybreaks[1]

\newcommand{\si}{\sigma}
\newcommand{\al}{\alpha}
\newcommand{\bt}{\beta}

\newcommand{\lam}{\lambda}
\newcommand{\ka}{\kappa}

\newcommand{\De}{\Delta}
\newcommand{\Ga}{\Gamma}

\newcommand{\bsig}{{\bm\sigma}}
\newcommand{\btau}{{\bm\tau}}

\newcommand{\be}{\begin{equation}}
\newcommand{\ee}{\end{equation}}

\newcommand{\bea}{\begin{eqnarray}}
\newcommand{\eea}{\end{eqnarray}}
\newcommand{\bd}{\begin{displaymath}}
\newcommand{\ed}{\end{displaymath}}
\newcommand{\ba}{\begin{array}}
\newcommand{\ea}{\end{array}}
\newcommand{\bi}{\begin{itemize}}
\newcommand{\ei}{\end{itemize}}
\newcommand{\bc}{\begin{center}}
\newcommand{\ec}{\end{center}}
\newcommand{\bfl}{\begin{flushleft}}
\newcommand{\efl}{\end{flushleft}}
\newcommand{\bfr}{\begin{flushright}}
\newcommand{\efr}{\end{flushright}}
\newcommand{\non}{\nonumber}

\newcommand{\bl}{\begin{aligned}}
\newcommand{\el}{\end{aligned}}

\newcommand{\hde}{\hat{\delta}}

\newcommand{\hchi}{\hat{\chi}}
\newcommand{\hxi}{\hat{\xi}}

\newcommand{\bJ}{\bar{J}}

\newcommand{\barxi}{\bar{\xi}}

\newcommand{\tE}{\tilde{E}}
\newcommand{\tM}{\tilde{M}}

\newcommand{\fs}{\frac{1}{2}}

\newcommand{\ra}{\rangle}
\newcommand{\la}{\langle}

  \def\bq{{\bf q}}

\def\bQ{{\bf Q}}   
\def\bH{{\bf H}} \def\bh{{\bf h}} 
 \def\bd{{\bf d}}  \def\bJ{{\bf J}}

 \def\bJ{{\bf J}}

\def\={\!\!\!&=&\!\!\!}
\def\+{\!\!\!&&\!\!\!+~}
\def\-{\!\!\!&&\!\!\!-~}

\begin{document}

\title{Anisotropic magnetic and quadrupolar H-T phase diagram of CeRh$_\text 2$As$_\text 2$}

\author{Burkhard Schmidt and Peter Thalmeier}
\affiliation{Max Planck Institute for Chemical Physics of Solids, 01187 Dresden, Germany}
\date{\today}

\begin{abstract}
The tetragonal heavy fermion compound CeRh$_2$As$_2$ has intriguing low temperature symmetry breaking phases whose nature is unclear. The unconventional superconducting phase is complemented by other normal state phases which presumably involve ordering of 4f electron multipoles supported by the Kramers doublets split by the tetragonal crystal electric field (CEF). The most striking aspect is the pronounced anisotropic $H$-$T$ phase boundary for in-plane and out-of plane field direction. Using a localized 4f CEF model we demonstrate that its essential features can be understood as the result of competing low field easy-plane magnetic order and field-induced quadrupolar order of XY type. We present calculations based on a coupled multipole random-phase approximation (RPA) response function approach as well as a molecular field treatment in the ordered regime. We use an analytical approach for a reduced quasi-quartet model and numerical calculations for the complete CEF level scheme. We discuss the quantum critical properties as function of multipolar control parameters and explain the origin of a pronounced a-c anisotropy of the $H$-$T$ phase 
diagram. Finally the field and temperature evolution of multipolar order parameters is derived and the high field phase diagram is predicted.
\end{abstract}

\maketitle

\section{Introduction}
\label{sec:introduction}

\begin{figure}
\label{fig:cryst}
\includegraphics[width=.55\columnwidth]{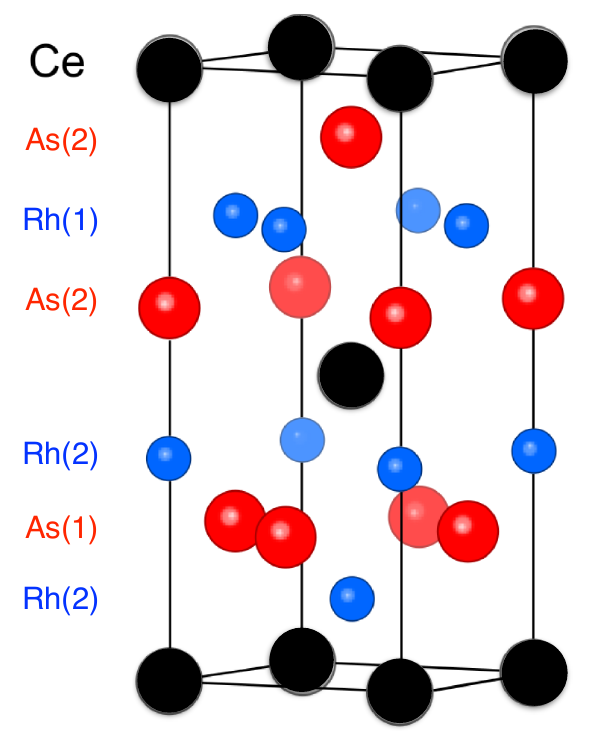}
\caption{Conventional unit cell of locally noncentrosymmetric CeRh$_2$As$_2$. The Ce site has $C_{4v}$ symmetry and is not an inversion center~\cite{fischer:23}, concomitantly two inequivalent Rh as well as As sites exist.}
\end{figure}

\begin{figure}
\includegraphics[width=.95\columnwidth]{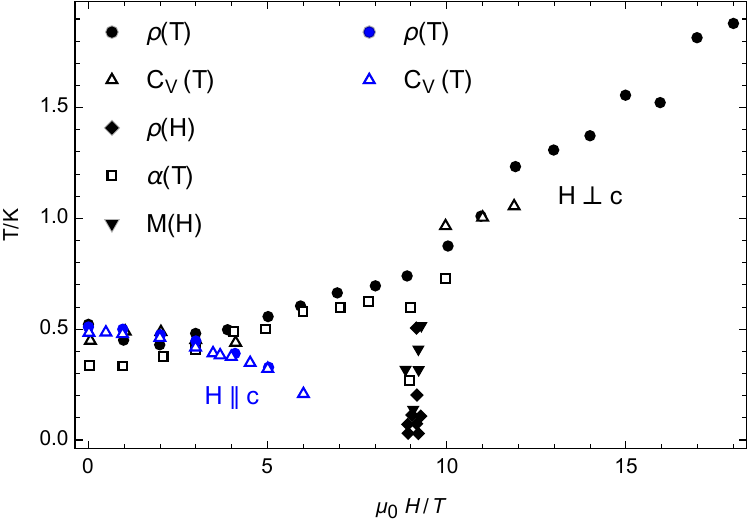}
\caption{Normal state $H$-$T$ phase boundaries for CeRh$_2$As$_2$ adapted from Hafner et al~\cite{hafner:22} for field direction $\perp c$ (black symbols) and Semeniuk et al.~\cite{semeniuk:23} for field direction $\parallel c$ (blue symbols). The corresponding experimental methods are indicated in the legend. Their appearance and anisotropy is in qualitative agreement with the calculated curves in Figs.~\ref{fig:tm-h0}(a) and~\ref{fig:tm-full}(a). For the comparison of typical field scales see Table~\ref{tbl:critval}.}
\label{fig:tcr-exp}
\end{figure}

The tetragonal (C$_{4\text v}$) compound CeRh$_2$As$_2$ has been added to the list of heavy fermion systems with complex symmetry breaking phases at low temperature and fields. Primarily superconductivity (SC) was discovered~\cite{khim:21} with $T_\text c=0.3\,\rm K$ and proposed~\cite{landaeta:22,ogata:23,semeniuk:23} to be of unconventional nature, in particular for field along the c-axis it was suggested that a transition between even and odd parity state takes place in an external field. This transition is claimed to be connected to the locally noncentrosymmetric crystal symmetry~\cite{fischer:23} as shown in Fig.~\ref{fig:cryst} and exemplified by the lack of local inversion symmetry at the f-electron sites. This has, however, negligible influence on the localized 4f electrons since CEF potentials which are odd under inversion have no effect within the f-electron subspace. Furthermore the SC phase was found to be surrounded by other phases which break the symmetry in the normal state~\cite{hafner:22,mishra:22,semeniuk:23,chajewski:24} with a zero-field $T_0=0.5\,\rm K$ which is a familiar scenario for heavy fermion compounds~\cite{thalmeier:04}. As for the SC phase what kind of order parameters are involved is still unidentified but $\mu$SR experiments suggests that spontaneous magnetic moments are formed below $T_0$ although their size is still unknown~\cite{khim:24}. The magnetic order has been excluded to be of FM type~\cite{hafner:22,semeniuk:23} but the ordering wave vector is so far not known, we may conjecture that it is of AF type. This may be drawn from calculated Fermi surface nesting properties~\cite{wu:24} and observed spin fluctuation peaks~\cite{chen:24a} at the AF wave vector $\bq=(\pi,\pi)$. Furthermore NMR experiments~\cite{kitagawa:22} indicate that magnetism should be of the easy-plane type, judging from spin fluctuations in the disordered phase. Like SC, this unidentified phase appearing below $T_0$ has, however, a striking a-c axis anisotropy of the phase boundary shown in Fig.~\ref{fig:tcr-exp} that is incompatible with conventional antiferromagnetism. It cannot be explained by a simple exchange anisotropy associated with the $C_{4v}$ CEF ground state doublet. Firstly the anisotropy is only moderate as expressed by similar dipolar ground state matrix elements in both directions. Secondly such anisotropy can never lead to the appearance of another high-field phase with {\it strongly increasing} ordering temperature as is observed for in-plane field in Fig.~\ref{fig:tcr-exp}. This requires the presence of additional degrees of freedom to be involved. A natural candidate are quadrupoles supported by the Ce$^{3+}$ ($J=5/2$) CEF states that consist of three Kramers doublets where the lower two may be considered to form a quasi-quartet. At the end of Sec.~\ref{sec:CEFmodel} we argue in detail which kind of quadrupole can be candidate.

Since CeRh$_2$As$_2$ is a heavy fermion system hybridization with conduction electrons is present. The estimated Kondo temperature $T^*$ (band width of heavy quasiparticles) is of the same order as the splitting $\De$ of the quasi-quartet system~\cite{hafner:22,christovam:24}. However when considering purely the question of symmetry breaking and the stability of multipolar phases the local 4f approach may be a reasonable starting point despite the presence of hybridization. This has been successfully demonstrated for the prominent pure quartet multipole order in CeB$_6$~\cite{shiina:97,thalmeier:21} and also in the quasi-quartet compound YbRu$_2$Ge$_2$~\cite{jeevan:06,takimoto:08} which show strong and moderate hybridization effects, respectively. The localized approach will also be used here for CeRh$_2$As$_2$ to investigate its most striking feature: the extreme anisotropy of the normal state phase diagram. In any case it is necessary to investigate its predictions as a reference point. Recent ARPES experiments~\cite{chen:24} have indeed proposed a predominantly localized character of 4f electrons in this compound and support such starting point.

For the determination of the $H$-$T$ phase diagrams we use the reduced quasi-quartet model in an analytical approach as well as the full CEF level scheme with three Kramers doublets in a numerical treatment. From symmetry arguments we identify in which configuration of conjectured dipolar and quadrupolar order parameters one may expect a strong a-c anisotropy of the $H$-$T$ phase diagram to appear. The most convenient technique of its determination is the response function formalism. Starting from the bare single-site CEF level susceptibilities we derive the coupled collective RPA multipole susceptibilities in the external field that include the intersite multipole interactions and follow their singularities from the disordered side which locates the phase boundaries. We show that the coupling of magnetic dipolar and quadrupolar moments happens through mixed multipole field-induced susceptibilities which appear only for the in-plane field direction. The field-induced mixing of the quasi-quartet doublets generates a quadrupolar ground state moment which through field induced coupling with the dipolar moment stabilizes the ordered phase for the in-plane field. This mechanism is absent for a field along c-axis, and this distinction lies at the origin of the observed anisotropy of the phase diagram which will be explained in a semi-quantitative way both within the quasi-quartet model and the full CEF level scheme.

Furthermore we show that at small intersite quadrupolar coupling the AF and field-induced quadrupolar (FIQ) are separated, the former appearing at small, the latter at larger fields. As the quadrupole interaction increases at a quantum critical point (QCP) the two phase boundaries approach and rapidly merge into a single phase boundary. We also calculated the field dependence of magnetic and quadrupolar order parameters to illustrate the change of character of the phase diagram as function of interaction control parameters. In addition we comment on the in-plane anisotropy of the critical fields and their dependence on control parameters. Finally we look at the very high field behavior and show that for the in-plane field the transition temperature of the mostly quadrupolar phase increases up to very large values and then drops steeply when the field strength becomes comparable to the quasi-quartet splitting.

\section{The CEF model for C\lowercase{e}R\lowercase{h}$_\text 2$A\lowercase{s}$_\text 2$ and its multipole moments}
\label{sec:CEFmodel}

The complete CEF Hamiltonian model comprising all three doublets appropriate for $J=5/2$ in C$_{4\text v}$ site symmetry has been discussed in Ref.~\onlinecite{hafner:22}. We extend the discussion and include the multipolar operators and the dependence of their matrix elements on the CEF mixing parameter $\theta$ as given in Appendix~\ref{sec:appCEF}. The latter has been determined from a fit to the high-temperature susceptibility of CeRh$_2$As$_2$. It turns out that the level sequence is $\Gamma_7^{(1)}$ ($0\,\rm K$, ground state); $\Gamma_6$ ($30\,\rm K$) and $\Gamma_7^{(2)}$ ($180\,\rm K$) where energies are given as equivalent temperatures ($k_\text B\equiv 1$). Therefore at moderate temperatures one has to deal only with a quasi-quartet system consisting of the lowest two doublets split by $\Delta=30\,\rm K$; this restriction is convenient for analytical calculations but fully numerical results comprising all three levels will also be presented. The wave functions of the quasi-quartet are denoted by $|\Gamma_{7\sigma}^{(1)}\ra\equiv |1\si\ra$ and $|\Gamma_{6\sigma}\ra\equiv |2\si\ra$. 

However, first we give a summary of all CEF properties for clarity and orientation. The Ce ions are located on a locally noncentrosymmetric lattice (space group $D_{4\text h}^7$ or P4/nmm) in layered tetragonal planes with site symmetry C$_{4\text v}$, highest rotational axis is a fourfold one. The formal charge is $3+$ and Hund's rules yield a ${}^2{\rm F}_{5/2}$ ground-state configuration. The $J=5/2$ CEF Hamiltonian, written in Steven's operator representation~\cite{hutchings:64,lea:62,kuramoto:09} is therefore
\begin{equation}
	{\cal H}_\text{CEF}
	=
	B_2^0O_2^0+B_4^0O_4^0+B_4^4O_4^4
	\label{eq:CEF}
\end{equation}
Its eigenvalues, the CEF level energies are obtained as
\begin{equation}
\begin{aligned}
	E_{\Gamma_7^{(1)}}
	&=
	4\left(B_2^0-15B_4^0\right)
	\\
	&\quad
	-6\sqrt{\left(B_2^0+20 B_4^0\right)^2+\left(2\sqrt5B_4^4\right)^2},
	\\
	E_{\Gamma_6}
 	&=
	-8 \left(B_2^0-15 B_4^0\right),
	\\
	E_{\Gamma_7^{(2)}}
	&=
	4\left(B_2^0-15B_4^0\right)
	\\
	&\quad
	+6\sqrt{\left(B_2^0+20 B_4^0\right)^2+\left(2\sqrt5B_4^4\right)^2},
\end{aligned}
\label{eq:CEFlevel}
\end{equation}
and the corresponding eigenstates consisting of 3 Kramers doublets are given in the basis of free ion states $|M\rangle$ ($|M|\leq\frac{5}{2}$) as
\begin{equation}
\begin{aligned}
	\left|\Gamma_7^{(1)}\right\rangle
	&=
	\cos\theta\left|\pm\frac52\right\rangle
	-\sin\theta\left|\mp\frac32\right\rangle,
	\\
	\left|\Gamma_6\right\rangle
	&=
	\left|\pm\frac12\right\rangle,
	\\
	\left|\Gamma_7^{(2)}\right\rangle
	&=
	\sin\theta\left|\pm\frac52\right\rangle
	+\cos\theta\left|\mp\frac32\right\rangle,
\end{aligned}
\label{eq:CEFstate}
\end{equation}
where $\theta$ is the mixing angle of the two $\Gamma_7^{(1),(2)}$ doublets that depends on all CEF parameters
according to	
\begin{equation}
	\theta
	=
	\frac12\tan^{-1}\left(\frac{2\sqrt5B_4^4}{B_2^0+20B_4^0}\right).
	\label{eq:CEFmixing}
\end{equation}
and its value is $\theta=0.346\pi$ for CeRh$_2$As$_2$. Therefore the relevant matrix elements of dipolar and quadrupolar operators also depend on this angle. They are listed in Table~\ref{tbl:matel} and plotted in Fig.~\ref{fig:matrix} of the Appendix~\ref{sec:appCEF}. (We note that our definition of $\theta$ differs to Hafner et al.~\cite{hafner:22} by setting $\theta\to\pi/2-\theta$ and we assume $B_4^4>0$.) The complete quadrupolar operator basis in the above CEF state space is explicitly given in Table~\ref{tbl:quadop}. In a large part of this work we will use the quasi-quartet model with two doublets split by $\Delta$. We stress that we use the proper wave functions for each doublet given above corresponding to the $\theta$-value for CeRh$_2$As$_2$ and not the wave functions of a fully degenerate cubic quartet.

These CEF states are unchanged in an external field $\bH_0$ parallel to the c-axis. However for $\bH_0 || a$ as described by the Zeeman term in Eq.~(\ref{eq:HAM}) or in a molecular field consisting of Zeeman term and internal polarization and order parameters (Eq.~(\ref{eq:MFOPa})) the bare CEF states given above are further mixed. Consequently the matrix elements of multipole operators depend on $\theta$ as well on the applied or molecular fields which have to be determined selfconsistently. This evaluation of matrix elements, expectation values and susceptibilities of the multipolar operators can be done either fully numerically or semi-analytically. The former case is necessary if we consider temperatures and fields whose effective energy scale is comparable to the lowest CEF splitting energy $\De$. The latter case is possible in the low-field and temperature range ($h,T<\De$) where one may restrict to the lowest quasi-quartet states as carried out in the main text. It is important to note that even though this inequality holds, the mixing of the excited $\Gamma_6$ into the ground state $\Gamma_7^{(1)}$ by the in-plane field is the essential mechanism for the appearance of a large anisotropy of phase boundaries because it induces a large quadrupole moment in the ground state as shown in Sec.~\ref{sec:qqmodel}.

\begin{table}
\caption{Quadrupole operator basis in C$_{4v}$ symmetry expressed in terms of angular momentum operators}
\label{tbl:quadop}
\[
\begin{array}{lll}
\Ga_1\;\;\; O_2^0 &= 3J_z^2-J(J+1),\\
\Ga_3\;\;\;	O_{x^2-y^2} &= J_x^2-J_y^2
	= \frac12\left(J_+^2+J_-^2\right),\\
\Ga_4\;\;\;	O_{xy} &= J_xJ_y+J_yJ_x
	= \frac1{2\rm i}\left(J_+^2-J_-^2\right),\\
\Ga_5^{(1)}\;	O_{yz} &= J_yJ_z+J_zJ_y \\
	&= \frac1{2\rm i}\left[
	\left(J_+-J_-\right)J_z+J_z\left(J_+-J_-\right)
	\right],\\
\Ga_5^{(2)}\; O_{zx} &= J_zJ_x+J_xJ_z \\
	&= \frac12\left[
	\left(J_++J_-\right)J_z+J_z\left(J_++J_-\right)
	\right].
\end{array}
\]
\end{table}
For compelling reasons discussed below Eq.~(\ref{eq:HAM}) we will restrict our model for the order parameters and phase diagrams in $\bH_0\parallel c,a$ to two candidates: The magnetic dipole $J_y$ (choosing the external field $\bH_0$ along x-axis) and the electric quadrupole $O_{xy}=(J_xJ_y+J_yJ_x)$ which break and preserve time reversal symmetry, respectively. This model might also be described in a pseudospin language~\cite{shiina:97,thalmeier:21} using $\bsig=\pm$ for the Kramers degree of each doublet and $\btau=1,2$ for the orbital degree of the two doublets. In order to avoid the various necessary state and operator mappings we here remain in the original basis of total angular momentum operators $\bJ$.

For the intersite interactions responsible for the possible broken symmetry phases we use the most rudimentary model containing magnetic out-of (c) and in-plane (a) nearest-neighbor (n.n.) exchange as well as a n.n. quadrupolar interaction. Since both ferromagnetic and ferro-quadrupolar orders are ruled out by experimental evidence the exchange is assumed to be of antiferro (AF) type for both multipoles. This conjecture is in agreement with the observed peak positions at $\bq=(\pi,\pi)$ in the effective RKKY interaction determined by the nesting vector of the calculated band structure of CeRh$_2$As$_2$~\cite{wu:24} and it also matches well with the peaks positions observed in inelastic neutron scattering~\cite{chen:24a}. Together with CEF potential and Zeeman term the model is then described by
\begin{equation}
\begin{aligned}
	H
	&=
	H_\text{CEF}
	-g_J\mu_\text B\mu_0\bH_0\cdot\sum_i\bJ_i
	\\
	&-\fs\sum_{\la ij\ra}J^c_{ij}J_i^zJ_j^z
	-\fs\sum_{\la ij\ra}J^a_{ij}\left(J_i^xJ_j^x +J_i^yJ_j^y\right)
	\\
	&-\fs\sum_{\la ij\ra}J^Q_{ij}O_{xy}(i)O_{xy}(j)
\end{aligned}
\label{eq:HAM}
\end{equation}
where we restricted to nearest-neighbor intersite interactions for the multipoles. Here $\bH_0$ with label $0$ always refers to the {\em external applied\/} field whereas later on fields with other labels or none at all ($\bH$) refer to the {\em internal molecular\/} fields that contain the effect of polarization and spontaneous order.

Since we restrict to n.n. $\la ij\ra$ sites within the tetragonal plane and the c-axis exchange is subdominant as seen from the susceptibility there are two interaction constants involved: i) the dipolar exchange constant $I_m=z|I_m^0|$ ($I^0_m<0$) where we suppress a possible $a,c$ exchange anisotropy, ii) the quadrupolar effective coupling $I_Q=z|I_Q^0|$ ($I^0_Q<0$) with $z=4$ denoting the n.n. coordination number. We also define the reduced external field as $\bh_0=g_J\mu_\text B\mu_0\bH_0$ and later likewise for the reduced molecular field $\bh$.

We now justify empirically and from symmetry arguments why we only include the $O_{xy}$ quadrupole interactions in the above Hamiltonian and the following analysis. The zero-field AF order is assumed to be of easy-plane type in agreement with the preceding high-temperature susceptibility anisotropy. Since $x$ and $y$ directions are equivalent we choose the ordered magnetic moment $J_y$. Then in a magnetic field along an axis $\alpha=z,x$, $H_0^\alpha J_y$ has to belong to the same irreducible representation as the quadrupole operator $O_\Gamma$ if there should be a mutual influence of $O_\Gamma$ and $J_y$. For $\alpha=z$ this would be $O_{yz}$ belonging to the doubly degenerate $O_{\Gamma_5}=(O_{yz},O_{zx})$. Since for this field direction a normal AF phase boundary is observed, there can be no induced quadrupole and therefore the $J^Q$ intersite interaction for $O_{\Gamma_5}$ must be negligible. For $\alpha=x$ it will be the component $O_{\Gamma_4}=O_{xy}$. In this field direction the phase diagram is strongly anomalous and therefore the $J^Q$ intersite interaction for $O_{\Gamma_4}$ should be important and included in Eq.~(\ref{eq:HAM}). The remaining $O_{x^2-y^2}$ and $O_2^0$ do not transform as $H_0^\alpha J_y$ for both field directions and therefore may be ignored. The latter, being the trivial representation already present in the CEF potential cannot act as an order parameter anyway. Therefore if we restrict to the two field directions mentioned the above model Hamiltonian is adequate. At the end of Sec.~\ref{sec:phasebound-qq} we will argue that for general in-plane direction of the field not treated here the $O_{x^2-y^2}$ quadrupole and its intersite coupling will have to be included as well.

\section{The coupled dipolar-quadrupolar RPA response functions}
\label{sec:dqresponse}

The $H$-$T$ multipolar phase boundaries which we intend to investigate are most conveniently determined by following the line of singularities for the collective RPA susceptibilities in the $H$-$T$ plane that marks the onset of long range order. It is also essential to obtain an understanding of the field and temperature dependence of the coexisting magnetic and quadrupolar order parameters inside the ordered region as may be obtained within the molecular field approximation (MFA) discussed in Sec.~\ref{sec:OP}. The determination of phase boundaries is much simpler with the former method and in fact will be carried out analytically within the quasi-quartet model. It is also possible but more time consuming within the MFA by solving the selfconsistent equation for the order parameters (Eq.~(\ref{eq:MFOP})) searching numerically for the temperature where they vanish. Within numerical accuracy both methods give the same result and we prefer the former involving the RPA susceptibilities in the para-phase.

For nonzero molecular field $\bH$ all CEF levels (Kramers doublets) are split leaving only singlets. In this case the static homogeneous single-site (non-interacting) response function for multipole operators $X_\al$ acting on the CEF states may be written as 
\begin{equation}
\begin{aligned}
	\chi^0_{\al\bt}(T,\bh)
	&=
	\sum_{n\neq m}\la n|X_\al |m\ra\la m|X_\bt |n\ra\frac{p_n-p_m}{E_m-E_n}
	\\
	&+
	\beta\left[
	\sum_n\la n|X_\al|n\ra\la n|X_\bt|n\ra p_n
	-\la X_\al\ra\la X_\bt\ra
	\right]
\end{aligned}
\label{eq:baresus}
\end{equation}
where the first and second terms are van Vleck and Curie contributions and $\beta=1/(k_\text B T)$. The energies $E_n$ and states $|n\ra$ are nondegenerate CEF eigenvalues and eigenstates in the molecular field $\bh=\bh_0-I_m\la J_{x,z}\ra$ and $p_n=Z^{-1}\exp(-\beta E_n)$ are their thermal occupations with $Z=\sum_n\exp(-\beta E_n)$ denoting the partition function. The notation $\langle X_\alpha\rangle$ denotes the thermal expectation value of the respective operator $X_\alpha$. In the dipolar molecular field $\bh$ the quadrupolar interaction appears only implicitly via the polarization $\la J_x\ra$. In the limit $\bh\rightarrow 0$ this will also lead to the correct form of the susceptibilities for the three degenerate Kramers doublets.

Now we include inter-site interactions of two different multipoles on nearest-neighbor sites $\la i,j\ra$ as defined by $I_{A,B}=\sum_{\la ij\ra}I_{A,B}(ij)$. Then the coupled $2\times2$ multipolar susceptibility matrix in $X_\al=(A,B)$ operator space is given by the RPA expression~\cite{jensen:91}
\begin{equation}
\underline{\chi}=[\underline{1}-\underline{I}\underline{\chi}^0]^{-1}\underline{\chi}^0
\label{eq:RPAsusmat}
\end{equation}
where $\underline{I}$ has only the diagonal matrix elements $I_A,I_B$. The diagonal susceptibility elements of the matrix $\underline{\chi}$ are then obtained as
\begin{widetext}
\bea
\chi_{AA}&=&\frac
{\chi^0_{AA}(1-I_B\chi^0_{BB})-I_A\chi^0_{AB}\chi^0_{BA}}
{1-I_A\chi^0_{AA}-I_B\chi^0_{BB}+I_AI_B(\chi^0_{AA}\chi^0_{BB}-\chi^0_{AB}\chi^0_{BA})}\non\\
\chi_{BB}&=&\frac
{\chi^0_{BB}(1-I_A\chi^0_{AA})-I_B\chi^0_{BA}\chi^0_{AB}}
{1-I_A\chi^0_{AA}-I_B\chi^0_{BB}+I_AI_B(\chi^0_{AA}\chi^0_{BB}-\chi^0_{BA}\chi^0_{AB})}\non\\
\label{eq:RPAsus}
\eea
\end{widetext}
where we denote $I_A=z|I_0^m|$ and $I_B=z|I_0^Q|$. Note that in a non-vanishing field $\bh$ the mixed multipole susceptibilities $\chi^0_{BA}=\chi^0_{BA}$ ($A\neq B$) are generally nonzero, depending on the symmetry of multipole operators. For the current model of three Kramers doublets in CeRh$_2$As$_2$ we consider the possible cases of coupled dipole ($\bJ$) and quadrupole ($O_{\Gamma}$) moments listed in Table~\ref{tbl:quadop} as incipient order parameters. This table identifies the dipoles and quadrupoles and their respective zero-field irreducible representations for point group C$_{4\text v}$ which can be mutually induced in a finite magnetic field along one of the tetragonal axes. As discussed above it shows that the most promising case of strong in-/out-of-plane anisotropy is the combination of a $\Gamma_4$ quadrupole and an in-plane magnetic $\Gamma_5$ dipolar order parameter on which we will focus in the following:

\noindent
 i) out-of-plane field $\bH_0=(0,0,H_0)$ and $A=J_y$ (or equivalently $J_x$) and $B=O_{xy}$;\\
ii) in-plane field $\bH_0=(H_0,0,0)$ and $A=J_y$ and $B=O_{xy}$.
 
\begin{table*}
\caption{Possible quadrupoles (see Table~\ref{tbl:quadop}) $O_\Gamma$ that can couple to dipoles $\bJ$ for symmetry field directions (in-plane $(H_x,H_y)$ or out-of-plane $H_z$). The $C_{4\text v}$ representation $O_\Gamma$ has to be contained in the product $\Gamma(\bH)\otimes \Gamma(\bJ)$. Note that~\cite{koster:63} $\Gamma_5\otimes\Gamma_5=\Gamma_1\oplus\Gamma_2\oplus\Gamma_3\oplus\Gamma_4$ (there is no $\Gamma_2$ quadrupole). The other products are trivial. We mention that the fully symmetric $O_2^0$ cannot be an order parameter since it does not break any local symmetry and therefore is already contained in the CEF Hamiltonian Eq.~(\ref{eq:CEF}) of the disordered phase. The case corresponding to the model in Eq.~(\ref{eq:HAM}) is given by the first line with $(H_x,0)\otimes(0,J_y)$ corresponding to a field-induced $O_{xy}$ quadrupole. For arbitrary in-plane field direction $O_{x^2-y^2}$ would have to be included.}
\label{tbl:symmetry}
\centering
\[
\setcounter{MaxMatrixCols}{11}
\begin{matrix}
\hline\hline
&\Gamma(\bH)\otimes \Gamma(\bJ) & \Gamma_1&\Gamma_3& \Gamma_4& \Gamma_5\\
\hline
\Gamma_5 \otimes \Gamma_5& (H_x,H_y)\otimes (J_ x,J_y) & - & O_{x^2-y^2}& O_{xy} & -& \\
\Gamma_5\otimes \Gamma_1& (H_x,H_y)\otimes J_z& - & -&-& (O_{yz},O_{zx}) \\
\hline
\Gamma_1\otimes \Gamma_5& H_z\otimes (J_x,J_y) &-& -& -& (O_{yz},O_{zx})\\
\Gamma_1\otimes \Gamma_1& H_z\otimes J_z & O_2^0& -&-&-\\
\hline\hline
\end{matrix}
\]
\end{table*}
In the first case i) we can read off from Table~\ref{tbl:symmetry} that the non-diagonal $\chi^0_{AB}$ vanishes identically. Then Eq.~(\ref{eq:RPAsus}) decouples to
\begin{equation}
\chi_{\al\al}=\chi^0_{\al\al}[1-I_\al\chi^0_{\al\al}]^{-1},\quad\al=A,B
\label{eq:RPA0}
\end{equation}
for the diagonal components and the field dependences for A, B are mutually independent. In the second case ii) of in-plane field the product $J_xH_y$ belongs to the same C$_{4\text v}$ representation $\Gamma_4$ as $O_{xy}$ and therefore there will be a non-vanishing non-diagonal field-induced susceptibility component $\chi^0_{BA}=\chi^0_{AB}$, consequently the full expressions in Eq.~(\ref{eq:RPAsus}) have to be used for the diagonal dipolar and quadrupolar RPA response functions $\chi_{J_y,J_y}$ and $\chi_{O_{xy},O_{xy}}$. 

\section{Restricted paramagnetic quasi-quartet model}
\label{sec:qqmodel}

For understanding the mechanism of dipolar and quadrupolar coexisting order and the associated anisotropic phase diagram it is essential to investigate a simplified model which may be treated analytically giving closed solutions for transition temperatures and critical fields. This is possible if we restrict to the lowest two Kramers doublets $\Gamma_7^{(1)}$ and $\Gamma_6$ forming a quasi-quartet split by $\Delta\simeq30\,\rm K$ in Eq.~(\ref{eq:CEFlevel}). In fact if we set the tetragonal CEF parameters like $B_2^0\to0$, $B_4^0\to-B_4$, and $B_4^4\to-5B_4$ in Eq.~(\ref{eq:CEF}), $\Delta\rightarrow 0$ and these two doublets would form the cubic $\Gamma_8$ quartet for $J=5/2$ with $\theta=\fs\tan^{-1}(\sqrt{5}/2)=0.134\pi$. As stressed before we use, however, the proper doublet wave functions $|\Gamma_7^{(1)}\ra$, $|\Gamma_6\ra$ for $\theta=0.346\pi$ corresponding to CeRh$_2$As$_2$.

Using this simplified model naturally implies that we also restrict to fields and temperatures of the order of the quasi-quartet splitting $\Delta=30\,\rm K$ where the upper third quartet at $180\,\rm K$ is not yet relevant. In fact we are mostly in the limit $T,h\ll\Delta$. This allows further simplifications: We can neglect the thermal population of the $\Gamma_6$ and for the discussion of the ordered phase may approximately eliminate the excited $\Gamma_6$ to lowest order in the molecular fields. Nevertheless the nondiagonal dipolar and quadrupolar matrix elements between $\Gamma_7^{(1)}$ and $\Gamma_6$ and the associated mixing of states for in-plane fields provide the essential mechanism to create the anisotropy of phase boundaries.

\subsection{Essentials of the quasi-quartet model in the field}

\begin{figure}
\includegraphics[width=.8\columnwidth]{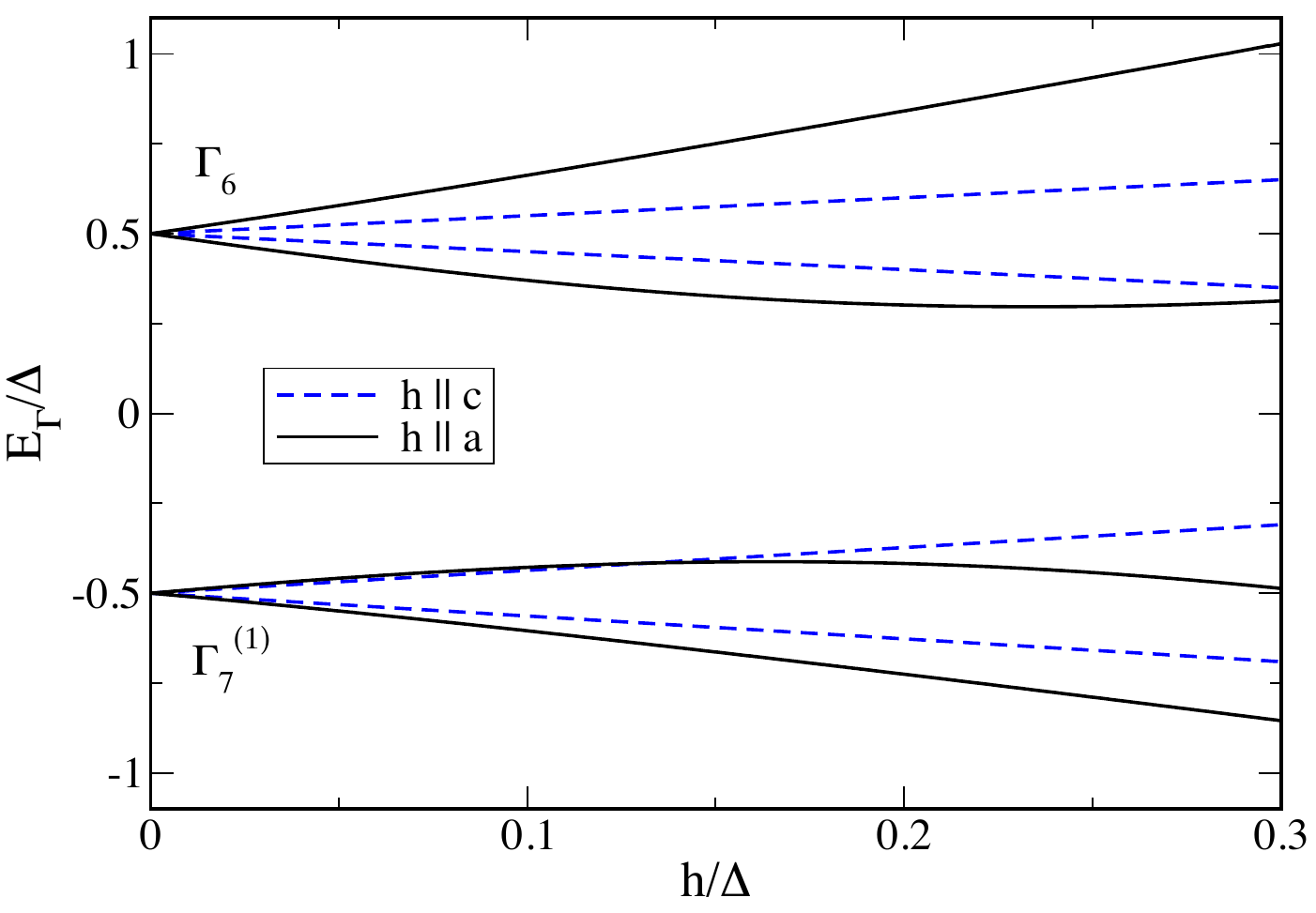}
\caption{Splitting of the quasi-quartet level scheme in the molecular field $h$ (equivalent to applied field $h_0$ for $I_m=0$) according to Eqs.~(\ref{eq:enc}) and~(\ref{eq:ena}). The relative size of the linear splitting of $\Gamma_7^{(1)}$ and $\Gamma_6$ is reversed when changing field direction from c to a in agreement with intra-doublet matrix elements in Table~\ref{tbl:matel}.}
\label{fig:qqlevel}
\end{figure}
To study the $H$-$T$ phase diagram we need to know the split CEF energies and corresponding eigenstates in an applied field. As described in detail in Sec.~\ref{sec:OP} it is, however, the molecular fields, generically called $\bH$ containing the applied field $\bH_0$ {\em and\/} the effect of polarization through intersite interactions that acts on each site. In terms of this effective molecular field $\bH$ the single site Hamiltonian is given by
\begin{equation}
H=\sum_\Ga\epsilon_\Ga|\Ga\ra\la\Ga|-g_J\mu_\text B\mu_0\bH\cdot\hat{\bf J}
\label{eq:HAM0}
\end{equation}
where $\epsilon_\Ga=(-\De/2,\De/2)$ and $|\Ga\ra$ refer to the (shifted) CEF energies and eigenstates of the quasi-quartet consisting of $|1\pm\ra$ and $|2\pm\ra$ (Eq.~\ref{eq:CEF}). The external field $\bH_0$ may be oriented parallel (${\bf H}_0=H_0\hat{{\bf z}}$) or perpendicular (${\bf H}_0=H_0\hat{{\bf x}}$) to the tetragonal plane, respectively. In the paramagnetic state it is aligned with the molecular field $\bh=\bh_0-I_m\la \bJ\ra$ where we defined $\bh=g_J\mu_\text B\mu_0\bH$ previously as the reduced field in equivalent energy units; later we will also use the dimensionless field strength $h'=h/\De$ normalized to the quasi-quartet splitting.

\paragraph*{Out-of-plane case $\bh\parallel \hat{{\bf z}}$ (c-axis).}
Due to the diagonal $J_z$ matrix the eigenstates will be unchanged but the Kramers doublet ($i=1,2$) energies split described by
\begin{equation}
E^\pm_{ci}=(-1)^i\frac{\De}{2}\mp m_{ci}h,
\quad
|E_i^{c\pm}\ra=|i\pm\ra.
\label{eq:enc}
\end{equation}
The matrix elements $m_{a,c\,i}^{(\prime)}$ used here and in the following are tabulated in Table~\ref{tbl:matel}.

\paragraph*{In-plane case $\bh\parallel \hat{{\bf x}}$ (a-axis).}
Now the $J_x$ matrix has nondiagonal elements (Eq.~\ref{eq:matdip}), therefore the CEF states will be mixed to new eigenstates. The corresponding mapping onto the new basis may be done by first performing a state rotation inside each doublet and then between the two doublets. The result for the four split level energies is ($i=1,2$):
\begin{equation}
\label{eq:ena}
\begin{aligned}
\tE^\pm_{ai}&=\cos^2\al_\pm E^\pm_{ai}+\sin^2\al_\pm E^\pm_{a\hat{\dotlessi}}
\\
&\phantom{=}
\pm(-1)^i\sin(2\al_\pm)(m'_ah)
\end{aligned}
\end{equation}
where analogous to Eq.~(\ref{eq:enc}) $E^\pm_{ai}=(-1)^i\frac{\De}{2}\pm m_{ai}h$ are the doublets split by their intrinsic linear Zeeman effect. We also used the notation $\hat{\dotlessi}=2,1$ for $i=(1,2)$. The mixing of the two levels is characterized by the angles $\al_\pm$ according to
\begin{equation}
\tan 2\al_\pm=\frac{\pm2m'_ah}{\De\pm(m_{a2}-m_{a1})h}
\to
\alpha_\pm\simeq\pm\frac{m'_ah}{\De}
\end{equation}
where the approximation holds for $h/\De\ll1$. The evolution of quasi-quartet level energies in the molecular field is shown in Fig.~\ref{fig:qqlevel} for the two field directions. Furthermore the eigenstates for the in-plane field are accordingly given by
\begin{equation}
\begin{aligned}
|\tE^\pm_{a1}\ra&=\cos\al_\pm|E^\pm_{a1}\ra-\sin\al_\pm|E^\pm_{a2}\ra,\\
|\tE^\pm_{a2}\ra&=\sin\al_\pm|E^\pm_{a1}\ra+\cos\al_\pm|E^\pm_{a2}\ra
\end{aligned}
\label{eq:eigen-a}
\end{equation}
with $|E^\pm_{ai}\ra=(|i+\ra\mp |i-\ra)/\sqrt{2}$ denoting the individual rotated Kramers doublet states in the transverse field.
 
For the calculation of the necessary response functions for $J_y$ and $O_{xy}$ one must transform the zero-field multipole operator matrices in Eqs.~(\ref{eq:matdip}) and~(\ref{eq:matqua}) (the $4\times4$ block) to the eigenstates in the applied field: i) For field in c-direction they are unchanged and $J_y$, $O_{xy}$ are identical to those in Eqs.~(\ref{eq:matdip}) and~(\ref{eq:matqua}). ii) For the a-direction ($\bH\parallel \hat{{\bf x}}$ chosen) the transformation to the new basis in Eq.~(\ref{eq:eigen-a}) leads to
\begin{align}
	J_x
	&=
	\begin{pmatrix}
		-M_{a1}^{+}& 0 &-M'_{a+}&0\\
		0&M_{a1}^{-}& 0& M'_{a-}\\
		-M'_{a+}&0&-M_{a2}^+&0\\
		0&M'_{a-}&0& M_{a2}^-\\
	\end{pmatrix},
	\label{eq:matfield-Jx}
	\\
	J_y
	&=
	{\rm i}
	\begin{pmatrix}
		0& -\tM_{a1} &0&\tM'_{a+}\\
		\tM_{a1}&0& -\tM'_{a-}& 0\\
		0&\tM'_{a-}&0& -\tM_{a2}\\
		-\tM'_{a+}&0&\tM_{a2} & 0\\
	\end{pmatrix},
	\label{eq:matfield-Jy}
	\\
	O_{xy}
	&=
	{\rm i}
	\begin{pmatrix}
		0&-\tM_Q&0&-\tM'_Q\\
		\tM_Q&0& -\tM'_Q&0\\
		0&\tM'_Q&0&-\tM_Q\\
		\tM'_Q&0&\tM_Q& 0\\
	\end{pmatrix}.
	\label{eq:matfield-O}
\end{align}
Comparing this with the c-axis field $\bH\parallel \hat{{\bf z}}$ direction in Eqs.~(\ref{eq:matdip}) and~(\ref{eq:matqua}) we notice that the essential difference are {\em field-induced\/} quadrupolar matrix elements $\tM_Q$ in the split ground state Kramers doublet $\Gamma_7^{(1)}$ which appear for the a-axis field orientation but are absent in the c-axis field direction (because the latter does not mix the CEF eigenstates). Therefore for the latter there will be no mutual dependence of dipole and quadrupole moments in the field whereas for the a-axis field direction such a dependence is induced. This distinction is at the origin of the strongly anisotropic behavior of phase boundaries in the two field directions as derived in detail below.

The other matrix elements for a-axis field are simply modified (or interchanged) as compared to the zero-field case. Explicitly we have for the dipolar operators (with $\hat{\dotlessi}=(2,1)$ for $i=(1,2))$:
\begin{equation}
\begin{aligned}
	M_{ai}^\pm
	&=
	m_{ai}\cos^2\al+m_{a\hat{\dotlessi}}\sin^2\al\mp m'_a\sin2\al,
	\\
	M'_{a\pm}
	&=
	m'_a\cos2\al\pm(1/2)(m_{a1}-m_{a2})\sin2\al,
	\\
	\tM_{ai}
	&=
	m_{ai}\cos^2\al-m_{a\hat{\dotlessi}}\sin^2\al,
	\\
	\tM'_{a\pm}
	&=
	m'_a\pm(1/2)(m_{a1}+m_{a2})\sin(2\al)
\end{aligned}
\label{eq:ymatel-a}
\end{equation}
and likewise for the quadrupolar operator 
\begin{equation}
\begin{aligned}
	\tM_Q
	&=
	m'_Q\sin(2\al),
	\\
	\tM'_Q
	&=
	m'_Q\cos(2\al).
\end{aligned}
\label{eq:Qmatel-a}
\end{equation}
Importantly the field induced $O_{xy}$ quadrupolar matrix element appears between the same split ground state wave functions as those of the dipolar $J_y$ operator enabling their coupling through mixed response functions.

One can see the origin of the induced matrix element directly in the low-field approximation where $\al\approx m'_ah/\De\ll\pi$. Then $\tM_\bQ\approx 2m'_am'_Qh/\De$ which shows that it is linear in h and proportional to {\em both\/} magnetic and quadrupolar matrix elements {\em between\/} the two doublets. Thus the transverse field mixes a $\Gamma_6$ component into the ground state with amplitude $\approx m'_ah/\De$ that forms an induced quadrupole ground state moment due to the $m'_Q$ non-diagonal original quadrupole matrix element $m'_Q$ between $\Gamma_7^{(1)}$ lowest and $\Gamma_6$ excited CEF Kramers doublet. This mechanism is decisive for the appearance of the high field induced quadrupolar phase.

\subsection{Response functions in the quasi-quartet system}
 
\begin{figure}
\includegraphics[width=.9\columnwidth]{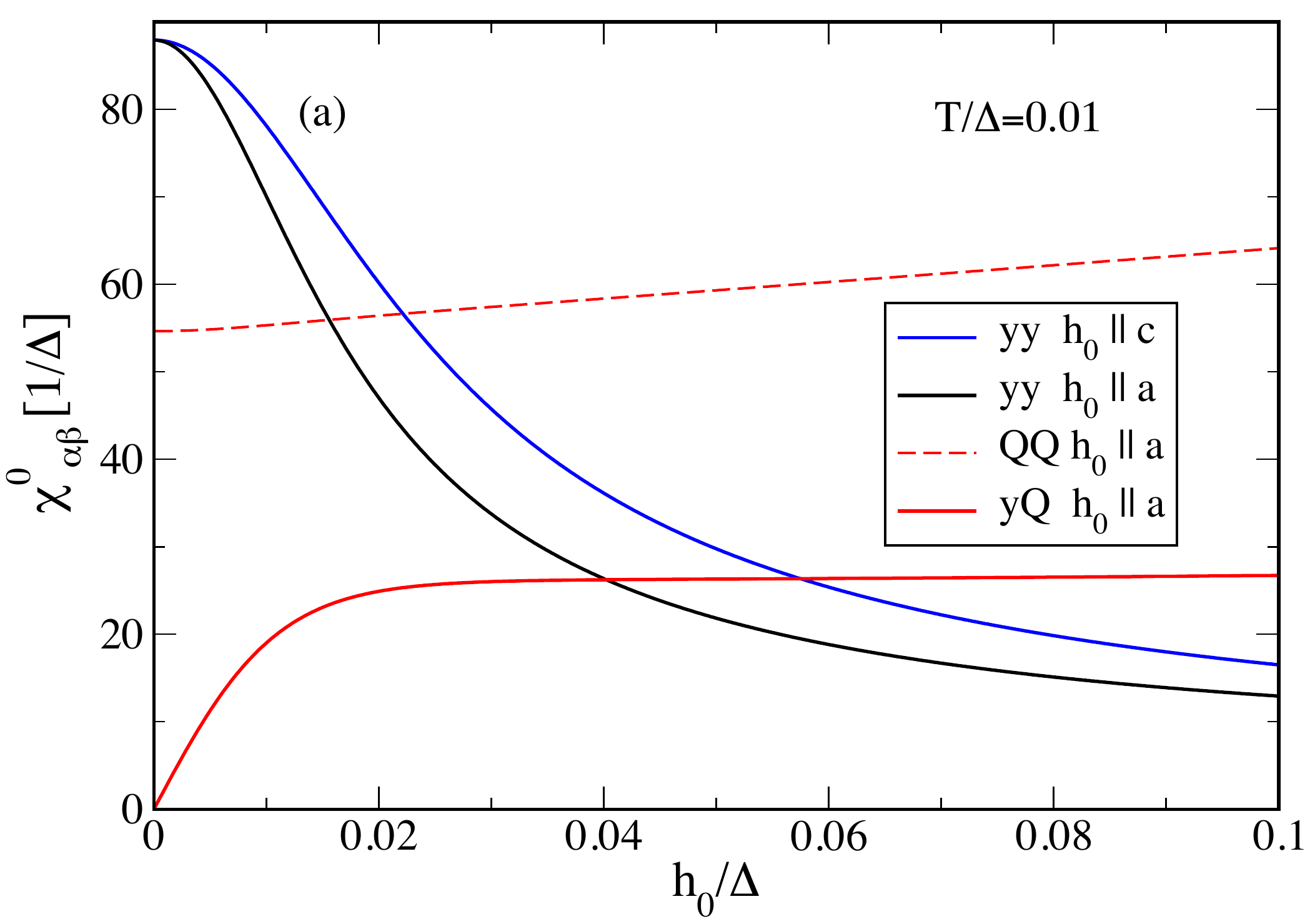}
\includegraphics[width=.9\columnwidth]{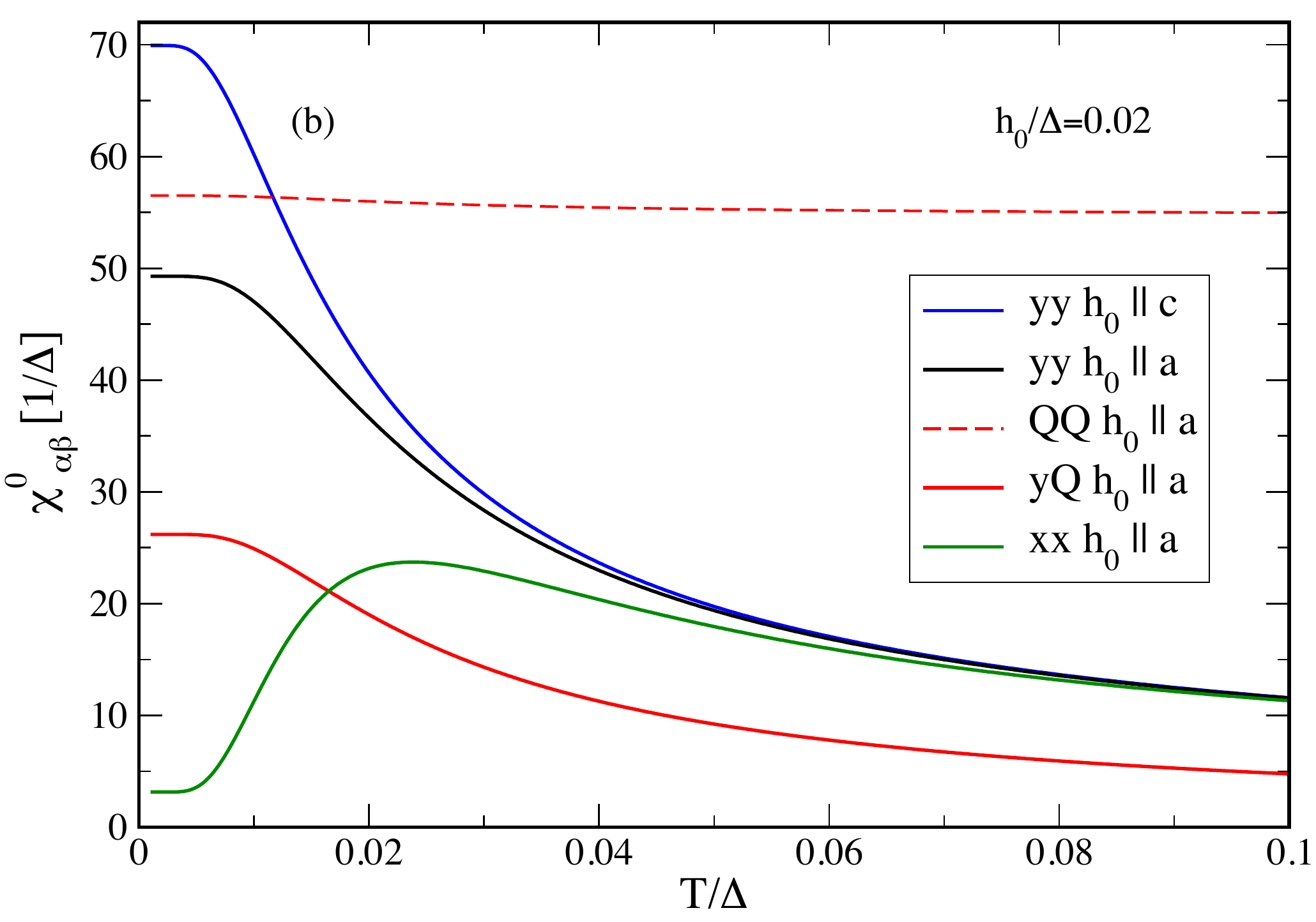}
\caption{Bare susceptibilities ($I_m=I_Q=0$) as function of applied external field $h_0$ and temperature. (a) The diagonal magnetic susceptibilities are suppressed in the field while the quadrupolar and mixed dipolar/quadrupolar 
$(\chi_{yQ}\equiv \chi_{Qy})$ ones increase with field. The latter is field-induced, vanishing for $h_0=0$. The crossing of $\chi_{yy}$ and $\chi_{yQ}$ appears in the region of the QCP in Fig.~\ref{fig:tm-h0}. (b) Susceptibilities dominated by pseudo-Curie ground state contribution show strong temperature dependence while the mostly van Vleck quadrupolar one shows very little $T$-dependence at low fields.}
\label{fig:sus0-h}
\end{figure}
With the above dipolar and quadrupolar matrices in the molecular field eigenstates we can now compute the bare multipolar response functions in Eq.~(\ref{eq:baresus}) that enter into the collective RPA susceptibilities in Eq.~(\ref{eq:RPAsus}). In this evaluation we assume that the effect of the splitting of upper levels and their thermal occupation may be neglected due to $h/\De, \;T/\De\ll1$ and the non-diagonal matrix elements are replaced by their zero-field values, independent of the field direction. However, the split ground state energies, occupations and matrix elements have to be treated exactly. Then we obtain

\paragraph*{out-of-plane case $\bh\parallel \hat{{\bf z}}$ (c-axis):}
\begin{equation}
 \chi^0_{yy}=\chi^0_{xx}=
 \frac{2m^{'2}_a}{\De}+2m_{a1}^2\frac{\tanh\frac{\hde_c}{2T}}{\hde_c}
\end{equation}
where $\hde_c=2m_{c1}h$ is the splitting of the $\Ga_7^{(1)}$ ground state doublet in the c-parallel field. As explained above for c-parallel field no induced quadrupole moment appears in the ground state and therefore the quadrupolar and mixed dipolar-quadrupolar susceptibilities are very small or vanish, respectively.

\paragraph*{in-plane case $\bh\parallel \hat{{\bf x}}$ (a-axis):}
\begin{equation}
\begin{aligned}
	\chi^0_{yy}
	&=
	\frac{2m^{'2}_a}{\De}
	+\frac{2\tM_{a1}^2}{\hde_a}\tanh\frac{\hde_a}{2T},
	\\
	\chi^0_{QQ}
	&=
	\frac{2m^{'2}_Q}{\De}
	+\frac{2\tM_Q^2}{\hde_a}\tanh\frac{\hde_a}{2T},
	\\
	\chi^0_{yQ}
	&=
	\left(\frac{2\tM_a\tM_Q}{\hde_a}+\frac{2m'_a m'_Q}{\De}\right)
	\tanh\frac{\hde_a}{2T} 
\end{aligned}
\label{eq:qqbaresus-a}
\end{equation}
where $\hde_a=2(m_{a1}\cos^2\al+m_{a2}\sin^2\al)h$ is the ground state doublet splitting in the a-parallel field. The dominant terms in these susceptibilities are the pseudo-Curie terms originating from the split ground state with a splitting energy $\hde_a$ and the matrix elements given in Eqs.~(\ref{eq:ymatel-a}) and~(\ref{eq:Qmatel-a}). The field dependence of these bare response functions is presented in Fig.~\ref{fig:sus0-h} which demonstrates the induced nature of the mixed response $\hchi^0_{yQ}$.

\subsection{Phase boundaries in the quasi-quartet model}
\label{sec:phasebound-qq}

The transition to the competing multipolar phases for $\bH_0 \parallel a$ appears when the RPA susceptibilities in Eq.~(\ref{eq:RPAsus}) diverge. This defines the phase boundary in the $H$-$T$ plane separating the para-phase form the ordered phase with non-vanishing magnetic and quadrupolar order parameters $\la J_y\ra$ and $\la O_{xy}\ra$ at each site (Sec.~\ref{sec:OP}). With $A=J_y$ and $B=O_{xy}$ the singularity appears if 
\begin{equation}
\begin{aligned}
	\mathop{\rm Det}\left(\underline1-\underline I\underline\chi^0\right)
	&=
	1-I_m\chi^0_{yy}-I_Q\chi^0_{QQ}
	\\
	&\quad
	+I_mI_Q\left(\chi^0_{yy}\chi^0_{QQ}-\chi^0_{yQ}\chi^0_{Qy}\right)=0
\end{aligned}
\label{eq:singularity}
\end{equation}
is satisfied, where $\chi^0_{yQ}=\chi^0_{Qy}$. Using Eq.~(\ref{eq:qqbaresus-a}) for the approximate bare susceptibilities this equation may be written in a more transparent form suitable for a closed solution for the critical temperature $T_\text{cr}(h)$ where the molecular field is given by $h=h_0-I_m\la J_x\ra$. For this purpose we introduce a set of appropriate dimensionless control parameters for dipolar as well as quadrupolar interactions characterized by the magnetic exchange $I_m\equiv I_A$ and quadrupolar $I_Q\equiv I_B$ effective interaction constants, respectively (see after Eqs.~(\ref{eq:HAM}) and~(\ref{eq:RPAsus})). They are given by
\begin{equation}
\begin{aligned}
	\xi_h^m
	&=
	\frac{2\tM_{a1}^2I_m}{\hde_a},\quad\xi_\De^m=\frac{2{m'}_a^2I_m}{\De},
	\\
	\xi_h^Q
	&=
	\frac{2\tM_Q^2I_Q}{\hde_a},\quad\xi_\De^Q=\frac{2{m'}_Q^2I_Q}{\De}.
\end{aligned}
\label{eqn:control}
\end{equation}
The $\xi_\De^{m,Q}$ are the control parameters for the non-diagonal van Vleck contributions to the susceptibility while the $\xi_h^{m,Q}$ are those for the pseudo Curie contributions associated with the split ground state doublet. The latter are the more important ones and strongly field dependent. For small fields $h/\De\ll1$ we have $\xi_h^m\sim 1/h$ due to the suppression caused by the splitting while $\xi_h^Q\sim h$ due to the induced quadrupole moment in the split ground state doublet. Hence for increasing field there is a tendency to suppress magnetic order in favor of induced quadrupolar order.

\begin{figure}
\includegraphics[width=0.80\columnwidth]{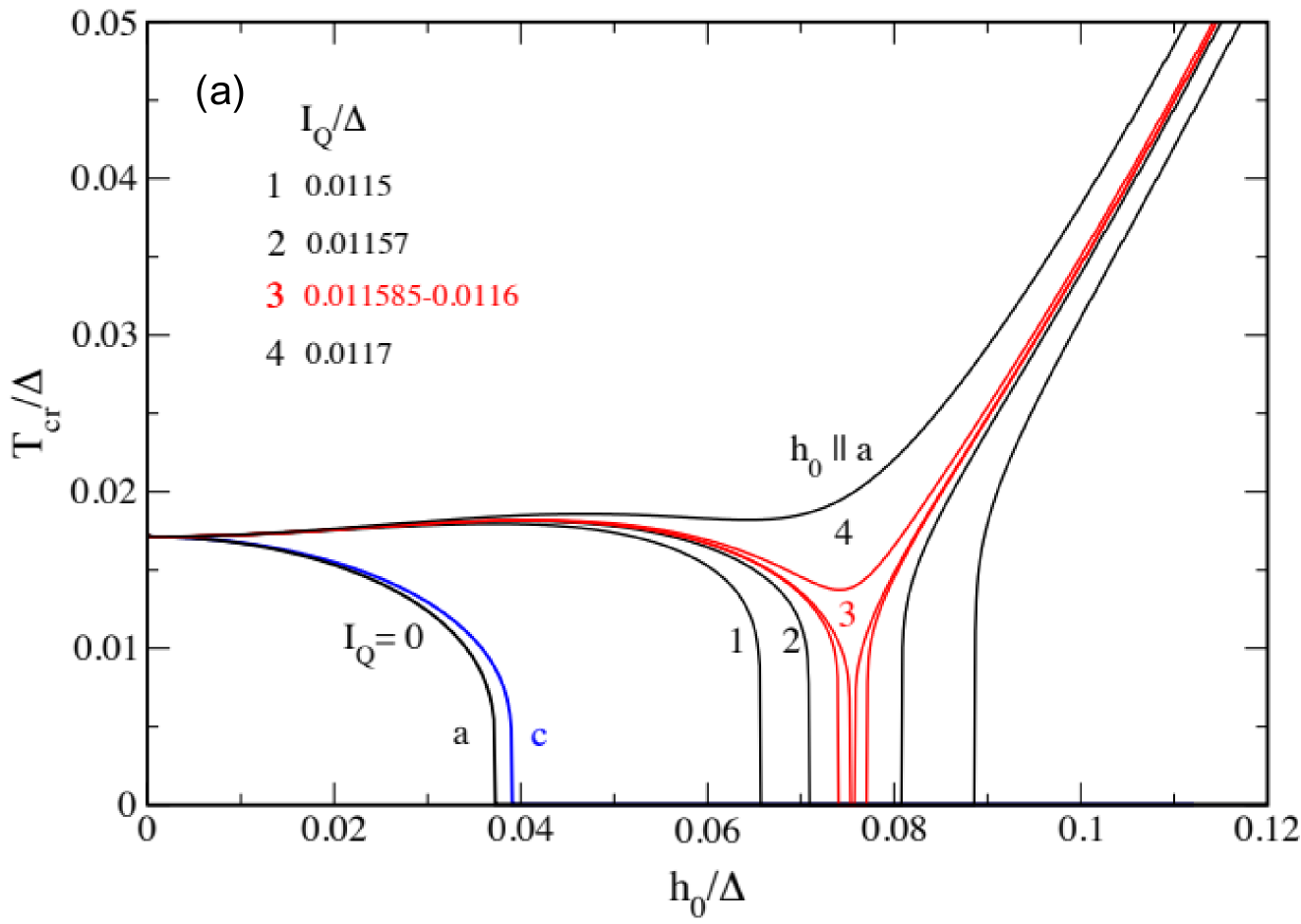}
\includegraphics[width=0.80\columnwidth]{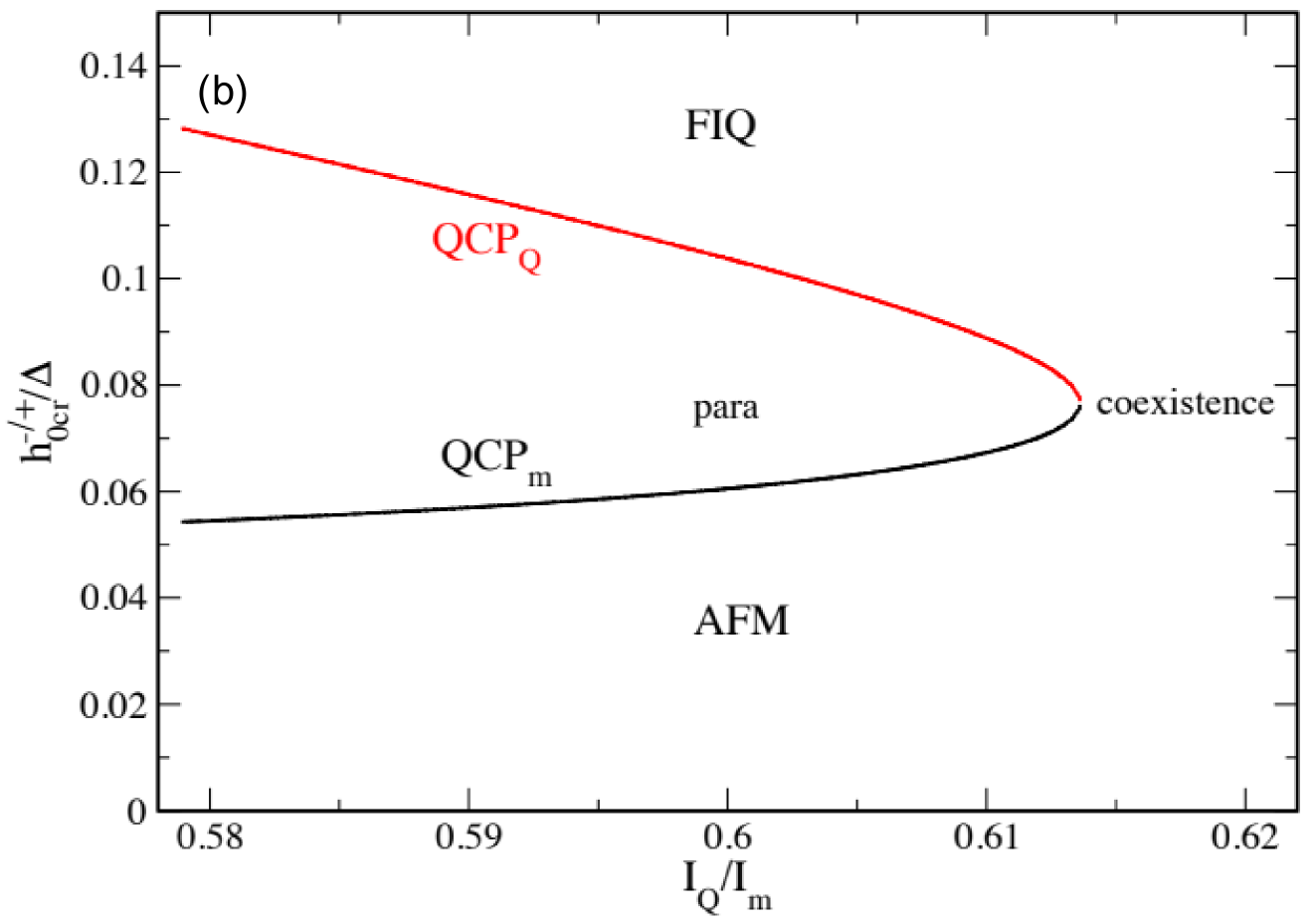}
\caption{(a) Low-field phase boundary curves $T_\text{cr}(h) =T_m$ (left) or $T_Q$ (right) for magnetic exchange $I_m=0.019$ and various quadrupolar coupling 
constants $I_Q$. Note that for $I_Q=0$ the critical field of AF order for a-direction would be slightly below the value for c-direction. For $I_Q > 0$, however, the former rapidly overtakes the latter while at the same time a new phase boundary appears at higher field that signifies the onset of field-induced quadrupolar (FIQ) order. The two phase boundaries approach each other for increasing $I_Q$ and touch at the QCP endpoint defined by Eq.~(\ref{eq:QCPend}). The red curves are for $I_Q$ close to the endpoint $I_Q\simeq0.0116$ where $(I_m,I_Q)$ correspond to the dimensionless control parameters $(\xi_\De^m,\xi_\De^q)\simeq(0.0596,0.633)$. For even larger $I_Q$ the two separate phase regions merge into one with coexisting AF and FIQ order (see also Fig.~\ref{fig:OP-h0})(b)) for all fields. (b) evolution of QCP lines $h^\pm_{0\text{cr}}$ for AFM and FQI as function of quadrupolar interaction strength, merging at the QCP endpoint $h_{0\text{cr}}$.}
\label{fig:tm-h0}
\end{figure}
With these expressions the singularity condition Eq.~(\ref{eq:singularity}) may be expressed as a quadratic equation for $\zeta_a=\tanh(\hde_a/2T)$:
\begin{equation}
\begin{aligned}
	&
	\barxi_\De(\barxi_\De+2\barxi_h)\zeta_a^2
	+(\rho_m\xi_h^Q+\rho_Q\xi_h^m)\zeta_a-\rho_m\rho_Q
	:=
	\\
	&
	A\zeta_a^2+B\zeta_a-C
	=0
\end{aligned}
\label{eq:boundary}
\end{equation}
where we used the abbreviations $\rho_m=1-\xi_\De^m$ and $\rho_Q=1-\xi_\De^Q$. Furthermore we defined the geometric means $\barxi_\De=(\xi_\De^m\xi_\De^Q)^\fs$ and $\barxi_h=(\xi_h^m\xi_h^Q)^\fs$. Then the solutions are $\zeta_a^\pm=[-B+\sqrt{B^2+4AC})]/(2A)$. Since $m_{a1}<0$ also $\hde<0$ and the physical solution is $\zeta^-_a\equiv\zeta_a$. The critical phase boundary is then finally given by
\begin{equation}
T_\text{cr}(h)=\frac{\hde_a(h)}{2\tanh^{-1}\zeta_a(h)}
\label{eq:tcr}
\end{equation}
The field dependence of the critical temperature is shown in detail in Fig.~\ref{fig:tm-h0}(a). Thereby the magnetic exchange $I_m$ has been fixed to reproduce the approximate experimental value $T_m(0)/\De=0.017$ and the curves are shown for different quadrupolar interaction parameters. It shows a separation into low field antiferromagnetic (AF) with transition temperature $T_\text{cr}\equiv T_m(h)$ and high-field field-induced quadrupolar (FIQ) phase (also of staggered type) with transition temperature $T_\text{cr}\equiv T_Q(h)$. A detailed discussion will be given in Sec.~\ref{sec:discussion}.

Here we want to further analyze the quantum critical point (QCP) and surrounding region where the two phases meet and merge into one with coexisting order parameters of both types throughout the whole field range. We are using the designation quantum critical in the sense that a zero-temperature phase change occurs as function of internal interaction control parameters $I_m,I_Q$ (equivalently $(\xi_\De^m,\xi_\De^Q)$) or external driving parameter (the applied magnetic field $h_0$) on which the ground state energy and hence the type of symmetry breaking depends. We are employing the RPA method in the para phase and a mean field treatment in the ordered phase. Therefore the influence of fluctuations caused by collective low energy modes in the ordered phase on the quantum critical properties (e.g. scaling behavior of the order parameter close to the QCP lines or endpoint) is naturally not included in our treatment. This would require a dynamical extension of Eq.~(\ref{eq:RPAsus}) for the coupled dipolar and quadrupolar degrees of freedom in the ordered state.

\begin{figure}
	\includegraphics[width=0.80\columnwidth]{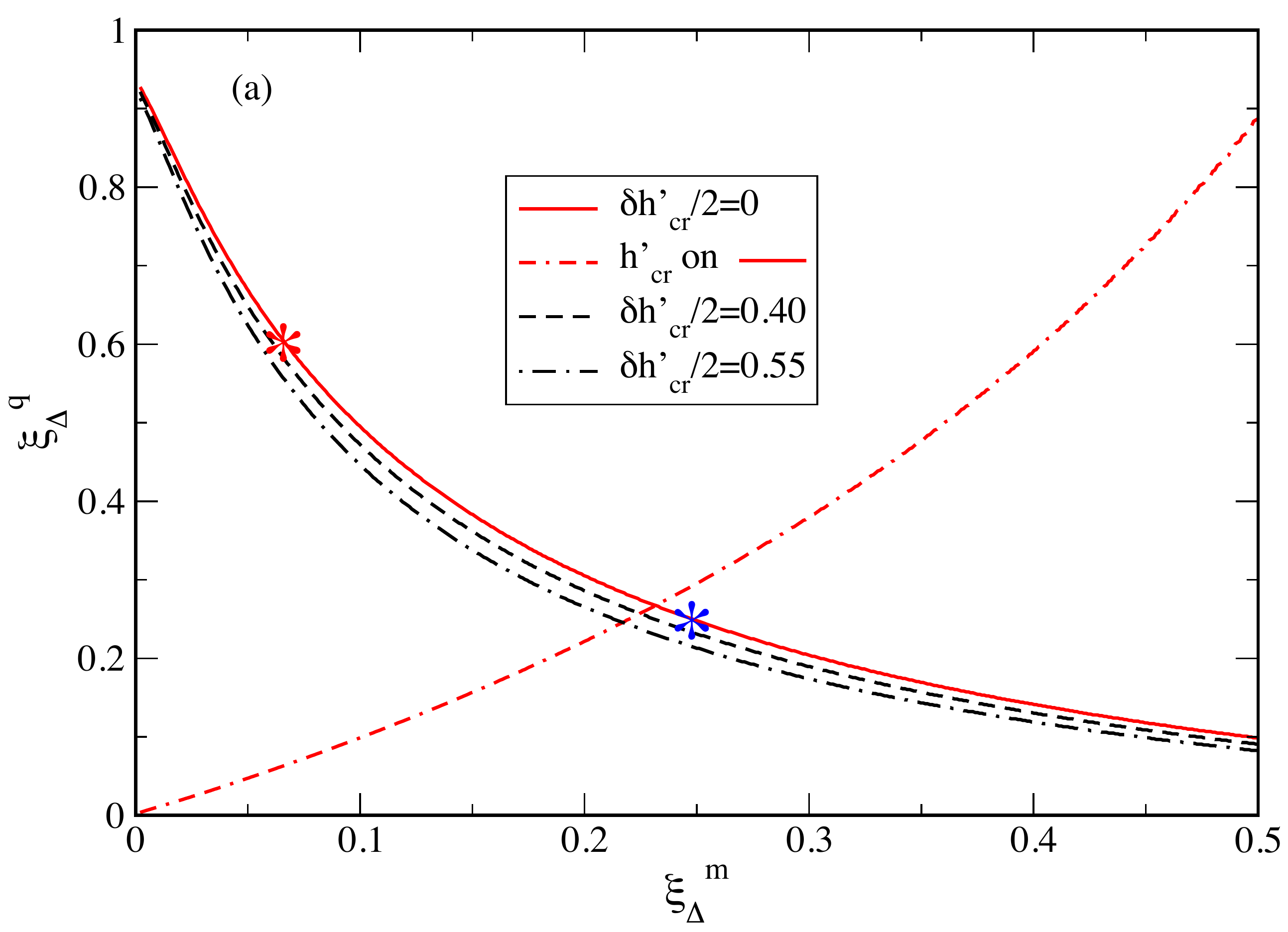}
	\includegraphics[width=0.80\columnwidth]{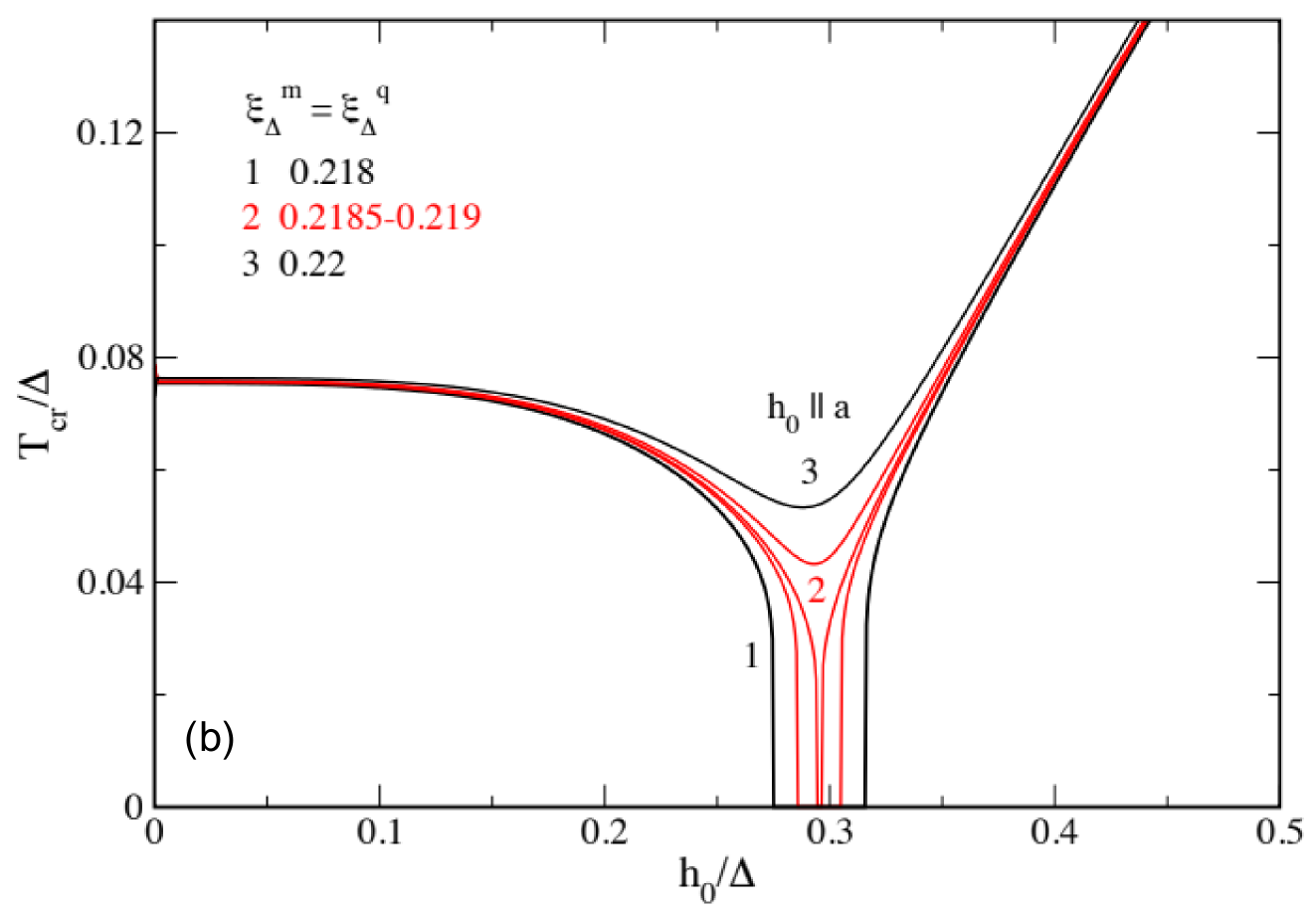}
	\caption{(a) Contours of constant $\delta h'_\text{cr}/2$ in the plane of dimensionless control parameters defined in Eq.~(\ref{eqn:control}). Along the full red line $\delta h'_\text{cr}/2=0$ the QCP's of AFM and FIQ phases merge to the critical endpoint on the red line. The red and blue stars correspond to the asymmetric interaction parameter case of Fig.~\ref{fig:tm-h0}(a), realized in CeRh$_2$As$_2$ and to the present (hypothetical) symmetric case in (b), respectively. The red dashed line gives the evolution of the critical field $h'_\text{cr}$ of the QCP endpoint (i.e.when moving on the red full line). Away from the latter the difference $\delta h'_\text{cr}$ of the two QCP$_{m,Q}$ critical fields increases rapidly (dashed and dash-dotted black lines) in accordance with Fig.~\ref{fig:tm-h0}(b). In (a) $h'_\text{cr}$ and $\delta h'_\text{cr}$ denote molecular fields as given in Eqs.~(\ref{eq:QCPend}) and~(\ref{eq:hcr}). (b) Critical temperatures as function of applied field for (hypothetical) symmetric case $\xi^m_\De=\xi_\De^q\simeq 0.22$, cf. Fig.~\ref{fig:tm-h0}(a) for the asymmetrical case of CeRh$_2$As$_2$.}
	\label{fig:hcr-xim}
\end{figure}
The zeroes of the transition temperature in Eq.~(\ref{eq:tcr}) are approached at the two quantum critical fields $h_\text{cr}^\pm$ when the denominator diverges, i.e. $\zeta_a(h_\text{cr}^\pm)\rightarrow -1$ (negative because $\hde <0$ due to m$_{a1}<0$). Their precise values may be read off from Fig.~\ref{fig:tm-h0}(a) obtained from Eqs.~(\ref{eq:boundary}) and~(\ref{eq:tcr}) but we also may derive closed expressions for the critical fields. Setting $\zeta_a=-1$ (corresponding to vanishing $T_\text{cr}$ in Eq.~(\ref{eq:boundary})) we arrive at the condition
\begin{equation}
2\barxi_\De^2-(\rho_m\xi_h^Q+\rho_Q\xi_h^m)+(\xi_\De^m+\xi_\De^Q)=1.
\label{eq:hcritcond}
\end{equation}
In lowest order in $h'=h/\De\ll1$ we find $\xi_h^m=\kappa\xi_\De^m/h'$ and $\xi_h^Q=\kappa^{-1}\xi_\De^Qh'$, therefore $\barxi_h=\barxi_\De$ independent of $h'$ to this order. Here we defined $\kappa=m_{a1}/(2{m'_a}^2)$ as a measure of the relative strength of diagonal and non-diagonal dipolar matrix elements (Eq.~(\ref{eq:matqua})). The resulting quadratic equation determines the critical fields of the two QCP's as
\begin{equation}
\begin{aligned}
	{h'}_\text{cr}^\pm
	&=
	\tilde{B}\pm[\tilde{B}^2-\tilde{C}]^\fs,
	\\
	\tilde{B}
	&=
	|\ka|\frac{\fs-(\barxi^2_\De+\xi^{ar}_\De)}{(1-\xi^m_\De)\xi^Q_\De},
	\quad
	\tilde{C}=\kappa^2\frac{(1-\xi^Q_\De)\xi^m_\De}{(1-\xi^m_\De)\xi^Q_\De}
\end{aligned}
\label{eq:hcrpm}
\end{equation}
where in addition to the geometric mean $\barxi_\De=(\xi^m_\De\xi^Q_\De)^\fs$ we also use the arithmetic mean $\xi^{ar}_\De=\fs(\xi^m_\De+\xi^Q_\De)$. From this equation we conclude that two distinct $h_\text{cr}^\pm$ (QCP$_\text Q$ and QCP$_\text m$ lines in Fig.~\ref{fig:tm-h0}(b)) exist for $\tilde{D}=\tilde{B}^2-\tilde{C} > 0$ and merge at a unique QCP endpoint for $\tilde{D}=0$. For $\tilde{D}<0$ the AF and FIQ phases coexist in the whole field range and $h_\text{cr}^\pm$ no longer appears. We define the {\em relative\/} critical field difference of the two phases by the ratio $\delta h'_\text{cr}/2=({h'}^+_\text{cr}-{h'}^-_\text{cr})/({h'}^+_\text{cr}+{h'}^-_\text{cr})$. Then the endpoint is determined by $\delta h'_\text{cr}/2=0$ where explicitly
\begin{equation}
\begin{aligned}
	\delta h'_\text{cr}/2
	&=
	\frac{|\kappa|}{(1-\xi_\De^m)\xi_\De^Q}\tilde{D}(\xi_\De^m,\xi_\De^Q)^\fs,
	\\
	\tilde{D}(\xi_\De^m,\xi_\De^Q)
	&=
	\left[(\barxi^2_\De+\xi^{av}_\De)-\fs\right]^2
	\\
	&\quad
	-(1-\xi^m_\De)(1-\xi^Q_\De)\barxi^2_\De.
\end{aligned}
\label{eq:QCPend}
\end{equation}
The condition $\delta h'_\text{cr}/2=0$ defines a quantum critical path in the interaction control parameter plane $(\xi^m_\De,\xi^Q_\De)$ along which the two QCP's have merged into a single $h'_\text{cr}$. Along this path its size is given by
\begin{equation}
h'_\text{cr}=|\ka|\left(\frac{\xi^m_\De(1-\xi^Q_\De)}{\xi^Q_\De(1-\xi^m_\De)}\right)^\fs.
\label{eq:hcr}
\end{equation}
Since all our discussion was limited to the small field region, i.e. also $h_\text{cr}\ll1$ the above formula is only valid when $\xi^m_\De$, $1-\xi^Q_\De$ are moderately small. It is nevertheless useful to discuss the limiting cases qualitatively. i) For $(\xi_\De^m,\xi_\De^Q)\rightarrow (0,1)$, $h'_\text{cr}\rightarrow 0$, the magnetic phase vanishes and is replaced by a self-induced quadrupolar phase already at zero field. This is possible because for $\xi_\De^Q >1$ the nondiagonal $m'_Q$ leads to a spontaneous quadrupole order already without field assistance. ii) For $(\xi_\De^m,\xi_\De^Q)\rightarrow (1,0)$ the small quadrupolar control parameter demands a very large field $h'_\text{cr}$ to reach the merging point with the magnetic transition. This behavior is shown in Fig~\ref{fig:hcr-xim}(a) and discussed further in Sec.~\ref{sec:discussion}.

We now comment on what one should expect for the in-plane anisotropy of the phase boundaries when the field is rotated perpendicular to c-axis. For a general field direction $\bh_0=(h_{0x},h_{0y})$ new complications arise: Firstly, two different quadrupoles $(O_{xy},O_{x^2-y^2})$ can be induced (Table~\ref{tbl:symmetry}) and secondly, together with the dipoles $(J_x,J_y)$ all four multipole operators $X_\alpha$ lead to a full bare susceptibility matrix $\underline{\chi}^0_{\alpha\beta}$ and likewise for the RPA susceptibility matrix $\underline{\chi}_{\alpha\beta}$. The resulting in-plane anisotropy will depend crucially on the matrix elements and inter-site coupling of the two quadrupoles $O_{xy}\;(m'_Q,I_Q)$ and  $O_{x^2-y^2}\;(\tilde{m}'_Q,I_{\tilde{Q}})$. It is interesting to consider an extreme case of in-plane anisotropy for $\bH_0\parallel(1,1,0)$ and $I_{\tilde{Q}}=0$. Then $O_{xy}$ and $O_{x^2-y^2}$ interchange their roles, i.e. $O_{xy}$ will not be induced and since $I_{\tilde{Q}}=0$, $O_{x^2-y^2}$ has no effect. In this case the lower critical field in Fig.~\ref{fig:tm-h0}(a) will be the same as in the a-case (for $I_Q=0$) and the upper one will not exist, i.e., we would recover the bare magnetic phase diagram for this diagonal field direction.

Finally, in contrast to the in-plane case, the phase boundary for the out-of-plane field $\bH_0\parallel c$ is much simpler to calculate because only the susceptibilities are decoupled according to Eq.~(\ref{eq:RPA0}). Then the singularity for $\chi_{\al\al}$ $(\al=x,y)$ simply leads to the magnetic transition temperature
\begin{equation}
T^c_{m}(h)=\frac{\hde_c(h)}{2\tanh^{-1}\frac{1}{\hxi_h^m}}
\end{equation}
with the ground state splitting now given by $\hde_c(h)=2m_{c1}h$ and $\hxi_h^m=\xi_h^m(1-\xi_\De^m)^{-1}$ and the c-parallel control parameter is now $\xi_h^m=2{m'}_a^2I_m/\hde_c$. Furthermore the molecular field is given by $h=h_0-I_m\la J_z\ra$. The resulting critical field is then ${h'}_\text{cr}=|m_{c1}| I_m$, formally the same as in the in-plane case $T_m^a$ for $I_Q=0$. In fact the two $T_m^{a,c}$ (for $I_Q=0$) are rather similar in Fig.~\ref{fig:tm-h0}(a).

\section{Description of the ordered AF and FIQ phases}
\label{sec:OP}

Now we turn to a discussion of the ordered phases as characterized by the temperature and field dependence of magnetic and quadrupolar moments using the MFA for the Hamiltonian in Eq.~(\ref{eq:HAM}). Naturally this neglects the influence of fluctuations due to low energy collective modes in the ordered regime. This may in particular change the scaling behavior of order parameter close to the quantum critical line and endpoint. These details are beyond the scope of our treatment where we focus on the global structure and anisotropy of the phase diagram for which the mean field treatment seems adequate. Thereby, in accordance with the n.n. interaction model we assume AF multipole order with sublattices $\lam=A,B$. The effective single-site Hamiltonian containing three molecular fields to be determined selfconsistently is given by (with $ E_0^\text{mf}(T,H)$ being a constant):
\begin{equation}
H_\text{MF}=\sum_{i\lam}H_\text{MF}^\lam(i)+E_0^\text{mf}.
\end{equation}
For {\em out-of-plane\/} field direction ($\bh_0\parallel c$) there can be no induced $O_{xy}$-type quadrupole and therefore only dipolar molecular fields are present leading to
\begin{equation}
H_\text{mf}^\lam(i)=
H_\text{CEF}(i)-\left[h_zJ_z(i)+h^\lam_yJ_y(i)\right]
\end{equation}
with molecular fields associated with homogeneous polarization $\la J_z\ra$ and staggered dipolar order parameter $\la J_y\ra_\lam$ given by (here $\lam=\pm 1$ for AF sublattices $\lam=A,B$):
\begin{equation}
\begin{aligned}
h_z&=h_0-I_m\la J_z\ra,\\
h_y^\lam &=\lam h_y,\;h_y=-I_m\la J_y\ra.
\end{aligned}
\label{eq:MFOPc}
\end{equation}
For {\em in-plane\/} field direction $(\bh_0\parallel a)$ we have a more complex situation with three molecular fields including
that of the induced quadrupole:
\begin{equation}
\begin{aligned}
	H_\text{mf}^\lam(i)
	&=
	H_\text{CEF}(i)+H_1(i),
	\\
	H_1(i)
	&=
	-\left[h_xJ_x(i)+h^\lam_yJ_y(i)+h_Q^\lam O_{xy}(i)\right]
\end{aligned}
\label{eq:HMF}
\end{equation}
where the molecular fields corresponding to the in-plane dipolar $\la J_{x,y}\ra$ and the induced quadrupolar $\la O_{xy}\ra$ are now given by
\begin{equation}
\begin{aligned}
	h_x
	&=
	h_0-I_m\la J_x\ra,
	\\
	h_y^\lam
	&=
	\lam h_y;\; h_y=-I_m\la J_y\ra,
	\\
	h_Q^\lam
	&=
	\lam h_Q;\; h_Q=-I_Q\la O_{xy}\ra.
\end{aligned}
\label{eq:MFOPa}
\end{equation}
We stress again that $h_0$ is the external field and fields with any other index or none at all ($h$) are molecular fields. It is those that are determined from the selfconsistency equations. Then calculating the polarizations $\la J_z\ra$ or $\la J_x\ra$ for the obtained molecular fields the external field corresponding to the selfconsistent set of molecular field may be obtained from the first of the equations in Eqs.~(\ref{eq:MFOPc}) and~(\ref{eq:MFOPa}).

From the eigenvalues $E_n$ and eigenstates $|n,\lam\ra$ of this Hamiltonian that depend on the three expectation values the latter have to be determined selfconsistently according to $\la A\ra_\lam=\sum_n p_n\la n\lam| A |n\lam\ra$. Now $p_n=Z^{-1}\exp(-E_n/T)$ are the occupation of (fully split) CEF levels in the molecular fields with $Z=\sum_n\exp(-E_n/T)$ denoting their MF partition function. Using the above equations the temperature and field dependence of $\la J_x\ra$, $\la J_y\ra_\lam=\lam \la J_y\ra$ and $\la O_{xy}\ra_\lam=\lam \la O_{xy}\ra$ may be calculated numerically using the full CEF level scheme. For the purpose of deeper understanding of field induced polarization and mutual competition of order parameters it is, however, useful to investigate again the quasi quartet model within an analytical approach for the ordered phases. We also note that the underlying $\la J_y\ra_\lam$ is staggered (even in the field) and the applied field is uniform therefore the induced quadrupolar $\la O_{xy}\ra_\lam$ will also be staggered within the approximations used as expressed in the last of Eq.~(\ref{eq:MFOP}).

\subsection{Order parameters, polarizations and effective molecular fields, effective operator treatment}

The staggered order parameters and homogeneous polarizations in  Eqs.~(\ref{eq:MFOPc}) and~(\ref{eq:MFOPa}) may be obtained analytically by restricting  to the quasi-quartet model in the limit $h, T \ll \De$. We focus mainly on the most interesting case where competing order parameters exist: For {\em in-plane\/} field direction ($\bh_0\parallel a$) the calculation is rather involved due to the presence of the induced quadrupolar order parameter $\la O_{xy}\ra$. In the quasi-quartet space the MF Hamiltonian is given explicitly by a $4\times 4$ matrix that has now entries at all places:
\begin{equation}
	H_\text{MF}^\lam
	=
	\begin{pmatrix}
	-\frac{\De}{2}& -m_{a1}h^\lam_- &im'_Qh_Q^\lam&-m'_ah_+^\lam
	\\
	-m_{a1}h^\lam_+&-\frac{\De}{2}& -m'_ah^\lam_-& -im'_Qh^\lam_Q
	\\
	-im_Qh^\lam_Q&-m'_ah^\lam_+&\frac{\De}{2}&-m_{a2}h^\lam_-
	\\
	-m'_ah^\lam_-&im'_Qh^\lam_Q& -m_{a2}h^\lam_+& \frac{\De}{2}
	\end{pmatrix}
\end{equation}
where we defined the complex MF expressions $h^\lam_\pm=h_x\pm i\lam h_y$. Unlike for the paramagnetic case of Eq.~(\ref{eq:HAM0}) the eigenvalues and -states of this MF Hamiltonian can no longer be obtained analytically. Therefore we resort to the effective operator technique~\cite{zeiger:73} where, due to $h/\De\ll1$ the effect of the upper $\Gamma_6$ doublet is eliminated and incorporated in an effective ground state Hamiltonian whose energies and eigenstates can be computed analytically and likewise the dressed matrix elements of multipole operators in the split ground state doublets are obtained. This procedure, based on Brillouin-Wigner perturbation theory, leads to
\begin{equation}
\begin{aligned}
	H^\lam_\text{eff}
	&=
	\begin{pmatrix}
	-\frac{\De^*}{2}& \frac{\hde_\lam}{2} \\
	\frac{\hde^*_\lam}{2}&-\frac{\De^*}{2}\\
	\end{pmatrix},
	\quad
	-\frac{\hde_\lam}{2}
	=
	\fs(\hde_1+i\lam\hde_2),
	\\
	\hde_1
	&=
	-2\left(m_{a1}h_x+\frac{2}{\De}m'_am'_Qh_yh_Q\right),
	\\
	\hde_2
	&=
	2\left(m_{a1}h_y+\frac{2}{\De}m'_am'_Qh_xh_Q\right).
\end{aligned}
\label{eq:HAMeff}
\end{equation}
The diagonal element $-\De^*/2$ is a renormalized level position that plays no role, The effective MF energy levels $E_n$ $(n=\pm)$, shifted by $\frac{\De^*}{2}$ of the split $\Gamma_7^{(1)}$ 
ground state are then given by
\begin{equation}
E_\pm=\mp\fs|\hde|=\mp\fs(\hde_1^2+\hde_2^2)^\fs
%\left[(m_{a1}h_x+\frac{2}{\De}m'_am'_Qh_yh_Q)^2+
%\lam(m_{a1}h_y+\frac{2}{\De}m'_am'_Qh_xh_Q)^2\right]^\fs
\label{eq:spliteff}
\end{equation}
independent of sublattice $\lam=A,B$. The corresponding eigenstates $|\psi_{n\lam}\ra$ $(n=\pm)$ in the basis of the unperturbed $|1\pm\ra$ doublet states are represented by the columns of the unitary matrix
\begin{equation}
	U_\lam^\dag
	=
	\frac{1}{\sqrt{2}}
	\begin{pmatrix}
	1& e^{i\lam\phi} \\
	-e^{-i\lam\phi}&1\\
	\end{pmatrix},
\quad
\tan\phi=\hde_2/\hde_1
\label{eq:Umat}
\end{equation}
Using the procedure described in Appendix \ref{sec:appeff} this allows to compute the MF expectation values of operators in the case $t,h\ll\De$ with the result given by
\begin{widetext}
\begin{equation}
\begin{aligned}
	\la J_x\ra
	&=
	\left[-m_{a1}\cos\phi+\frac{2}{\De}m'_am'_Qh_Q\sin\phi\right]
	\tanh\frac{|\hde|}{2T}
	+\frac{2}{\De}{m'_a}^2h_x,
	\\
	\la J_y\ra_\lam
	&=
	-\lam\left[-m_{a1}\sin\phi+\frac{2}{\De}m'_am'_Qh_Q\cos\phi\right]
	\tanh\frac{|\hde|}{2T}
	+\lam\frac{2}{\De}{m'_a}^2h_y,
	\\
	\la O_{xy}\ra_\lam
	&=\lam\left[\frac{2}{\De}m'_am'_Q(h_x\sin\phi-h_y\cos\phi)\right]
	\tanh\frac{|\hde|}{2T}
	+\lam\frac{2}{\De}{m'_Q}^2h_Q.
\end{aligned}
\label{eq:MFOP}
\end{equation}
\end{widetext}
With the molecular fields given in Eq.~(\ref{eq:MFOPa}) this closed set of equations for the homogeneous polarization and the two staggered order parameters then has to be solved numerically. They have been written in a form to make their physical content transparent: i) In each operator expectation value the last term is due to the direct admixture of the the excited $\Gamma_6$ into the ground state $\Gamma_7^{(1)}$ by the molecular field connected with the {\em same\/} operator. ii) For $\la J_x\ra$ and $\la J_y\ra$ the first terms in parentheses are due to the direct contribution of the $\Gamma_7^{(1)}$ ground state to the expectation value. There is no such term for the quadrupolar $\la O_{xy}\ra$ order parameter since this operator has no (bare) matrix elements in the Kramers doublet ground state (Eq.~(\ref{eq:matqua})). iii) For $\la J_x\ra$ and $\la J_y\ra$ the second and for $\la O_{xy}\ra$ the only term in parentheses are due to induced matrix elements in the $\Gamma_7^{(1)}$ ground state caused by the {\em complementary\/} molecular field, i.\,e. quadrupolar $h_Q$ for the former and dipolar $h_x,h_y$ for the latter. This term leads to the mutual influence and competition of order parameters. It is also useful to check the purely magnetic case setting $I_Q=h_Q=0$. Then we obtain
\begin{equation}
\begin{aligned}
	\la J_x\ra
	&=
	-m_{a1}\cos\phi \tanh\frac{\hde}{2T}+\frac{2}{\De}{m'_a}^2h_x,
	\\
	\la J_y\ra_\lam
	&=
	\lam m_{a1}\sin\phi \tanh\frac{\hde}{2T} +\frac{2}{\De}{m'_a}^2h_y,
\end{aligned}
\end{equation}
where now we have the simplified $|\hde|=2|m_{a1}|(h_x^2+h_y^2)^\fs$ and $\tan\phi=-(h_y/h_x)$ without the $\sim h_Q$ contributions.

In the paramagnetic state with $\phi=0$ and $\la J_y\ra_\lam=0$ only the homogeneous polarization survives and is given by
\begin{equation}
\begin{aligned}
	\bh_0\parallel a: \la J_x\ra
	&=
	m_{a1}\tanh\frac{2m_{a1}h_x}{2T}+\frac{2}{\De}{m'}_a^2h_x,
	\\
	&\rightarrow
	|m_{a1}|+\frac{2{m'}^2_a}{\De}h_a,
	\\
	\bh_0\parallel c: \la J_z\ra
	&=
	m_{c1}\tanh\frac{2m_{c1}h_z}{2T}
	\rightarrow |m_{c1}|,
\end{aligned}
\label{eq:polpara}
\end{equation}
and the corresponding molecular fields are given by $h_{x,z}=h_0-I_m\la J_{x,z}\ra$. Here the arrows imply
the zero-temperature limit.

\section{Discussion of results for the quasi-quartet model}
\label{sec:discussion}

We start our discussion with the field and temperature behavior of the bare susceptibilities $\chi^0_{\al\beta}$ without intersite interactions shown in Fig.~\ref{fig:sus0-h}. The rapid reduction of dipolar components in the field (a) is due to the ground state splitting which suppresses their dominant Curie terms. The reduction is stronger for a- than for c- direction because $|m_{a1}|>|m_{c1}|$ (Table~\ref{tbl:matel}). The quadrupolar susceptibility which has no ground-state contribution for zero field is almost constant due to the dominant van Vleck term controlled by $m'_Q$. The most important aspect is the rapid field-induced increase of mixed $\chi^0_{yQ}$ susceptibilities which are allowed by symmetry, in contrast to $\chi^0_{xQ}$ which remains zero. The decrease of dipolar components and increase of the mixed $\chi^0_{yQ}$ suppresses the magnetic and favors the quadrupolar instability obtained from the interacting RPA susceptibilities in Eqs.~(\ref{eq:RPAsusmat}) and~(\ref{eq:RPAsus}). The complementary temperature dependence is shown in (b) with the expected decrease caused by the reduction of thermal population difference in the split ground state doublet. While $yy,zz$ remain finite the $xx$ component is reduced to zero at low temperatures; the latter plays no role in the ordering instabilities.

The location of instabilities defines the $H$-$T$ phase diagram of the model for CeRh$_2$As$_2$, i.e. the phase boundaries $T_\text{cr}(h_0)$ which are presented in Fig.~\ref{fig:tm-h0} based on the analytical calculation (Eq.~(\ref{eq:tcr})) for the quasi-quartet model. The magnetic exchange coupling $I_m$ is fixed such that $T_\text{cr}(0)/\De=0.017$ corresponding to the experimental value~\cite{hafner:22}.For absent quadrupolar coupling the magnetic $T_\text{cr}(h_0)=T_m(h_0)$ transition temperature behaves quite similar for both field directions. For c-direction it is slightly larger than for a-direction because as explained above the bare susceptibilities are also slightly larger for the former case.

The near {\em a-c isotropy\/} of the magnetic phase diagram for $I_Q=0$ means that the observed~\cite{hafner:22,semeniuk:23} pronounced {\em a-c anisotropy\/} and strong $T_\text{cr}(h_0)$ increase for large fields in CeRh$_2$As$_2$ demands the inclusion of other multipoles and their interaction beyond the purely magnetic dipoles. There are many examples in 4f compounds where this has also been observed like, e.g. Rare-Earth hexaborides~\cite{thalmeier:21}, 4f-skutterudites~\cite{kuramoto:09} and Yb-compounds~\cite{jeevan:06,takimoto:08,rosenberg:19}. As we have argued before the field induced coupling to the $O_{xy}$ quadrupole with its strong non-diagonal matrix element and resulting large bare and field-induced susceptibilities (Fig.~\ref{fig:sus0-h}) is a prime candidate. The effect of this inclusion on the critical field curves as function of the $O_{xy}$ intersite coupling strength is immediately seen in Fig.~\ref{fig:tm-h0}(a) as a strong {\em increase\/} of the (lower, AF) critical field $h^-_{0\text{cr}}$ with $I_Q$ and the concomitant appearance of a second transition at higher critical field $h^+_{0\text{cr}}$ which {\em decreases\/} with increasing $I_Q$ and represents a phase with primary quadrupolar order. For $h^-_{0\text{cr}}< h<h^-_{0\text{cr}}$ one has again a sector with fully disordered phase. Since the two values $h^\pm_{0\text{cr}}$ characterizing the QCP$_{m,Q}$ for the two order parameters move into opposite directions with increasing $I_Q$ this means that at a critical value of $I_Q$ the two critical field curves will touch and merge into one curve (the red curves in Fig.~\ref{fig:tm-h0}(a)) , i.e., the disordered sector vanishes at a QCP endpoint and one has coexisting AF and quadrupolar order throughout the field range $h\ll\De$ where the quasi-quartet model is applicable. Actually this does not change qualitatively when performing the numerical calculations for the full model comprising all CEF states as discussed below.

The opposite movement of the QCP$_{m,Q}$ fields with quadrupolar coupling is presented separately in Fig.~\ref{fig:tm-h0}(b) and it clearly demonstrates the merging in a quantum critical endpoint at around $I_Q\simeq 0.0116$. The real phase diagram in CeRh$_2$As$_2$ is qualitatively well described by our theoretical results close to the QCP endpoint (Fig.~\ref{fig:tcr-exp}). An observed phase line with almost constant $T_\text{cr}=T_{m}$, intercepted by a perpendicular phase boundary and after this a strong, almost linear increase of $T_\text{cr}=T_Q$ with field. The behavior of the magnetic and quadrupolar order parameters in the various sectors of the phase diagram will be discussed below.

It is worthwhile to avert the discussion of CeRh$_2$As$_2$ for a moment in favor of a more general perspective. It is interesting to follow the structure of the phase diagram and its segmentation as function of {\em both\/} interaction parameters, magnetic $I_m$ as well as quadrupolar $I_Q$ or better in terms of their associated dimensionless control parameters $\xi_\De^m$ and $\xi_\De^Q$. This is presented in Fig.~\ref{fig:hcr-xim}(a). It shows the line of QCP endpoints in the plane of control parameters that separates the coexistence phase with merged critical field lines from the region where two separate QCP$_m$ and QCP$_Q$ still exist. The contours correspond to the magnitude of the critical field splitting given in Eq.~(\ref{eq:QCPend}). On the (full) QCP endpoint line the asymmetric values $(\xi_\De^m, \xi_\De^Q) \simeq (0.059,0.633)$ correspond approximately to CeRh$_2$As$_2$. But the same qualitative phase diagram with touching critical field curves would be obtained with more symmetric control parameters $(\xi_\De^m, \xi_\De^Q) \simeq (0.219,0.219)$ as shown in Fig.~\ref{fig:hcr-xim}(b). In this case, however, the size of $T_\text{cr}(0)$ and $h_{0\text{cr}}$ with respect to CEF splitting $\De$ have increased by a significant factor as compared to the asymmetric case of CeRh$_2$As$_2$ (Fig.~\ref{fig:tm-h0}(a)).

\begin{figure}
	\includegraphics[width=0.80\columnwidth]{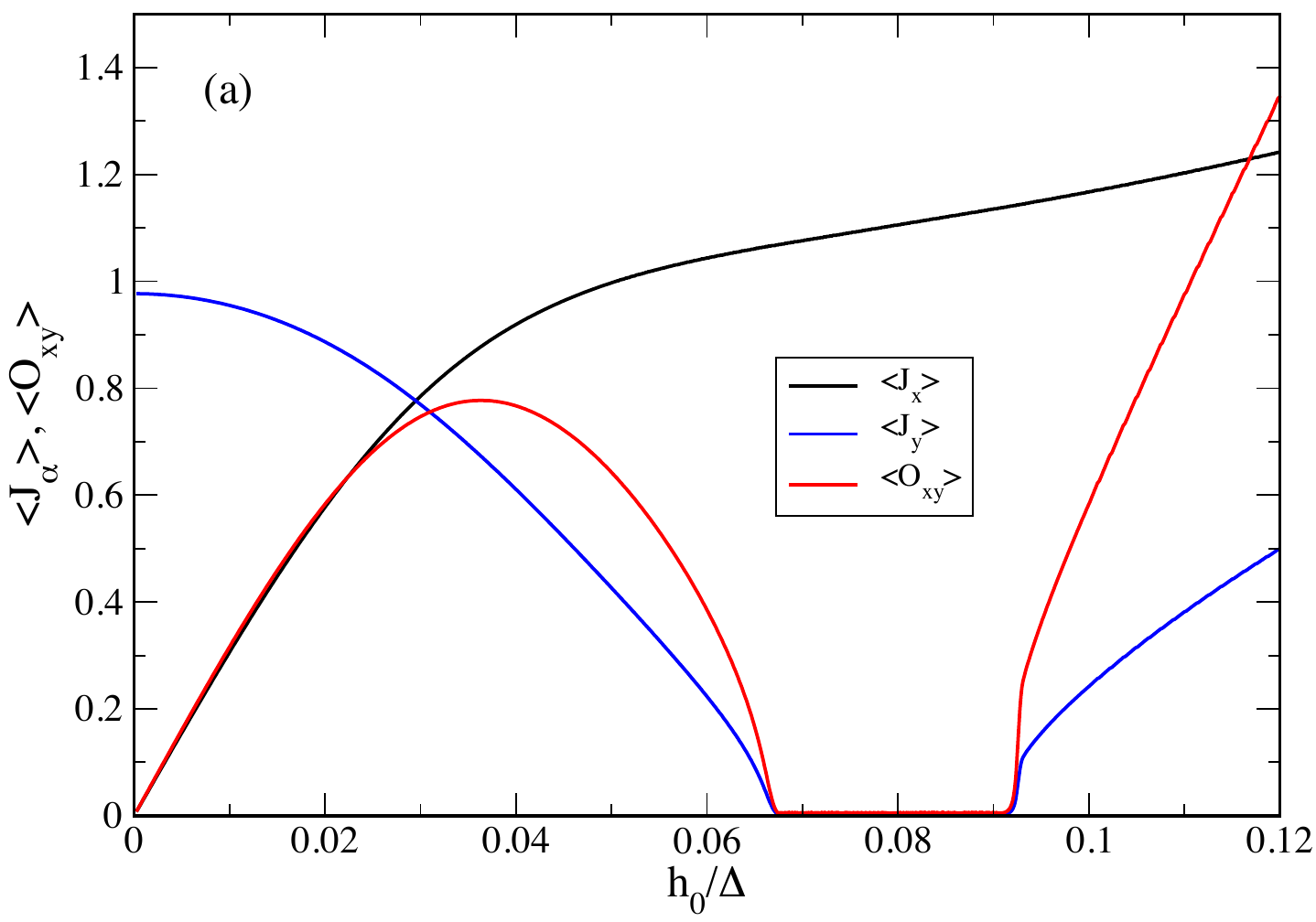}
	\includegraphics[width=0.80\columnwidth]{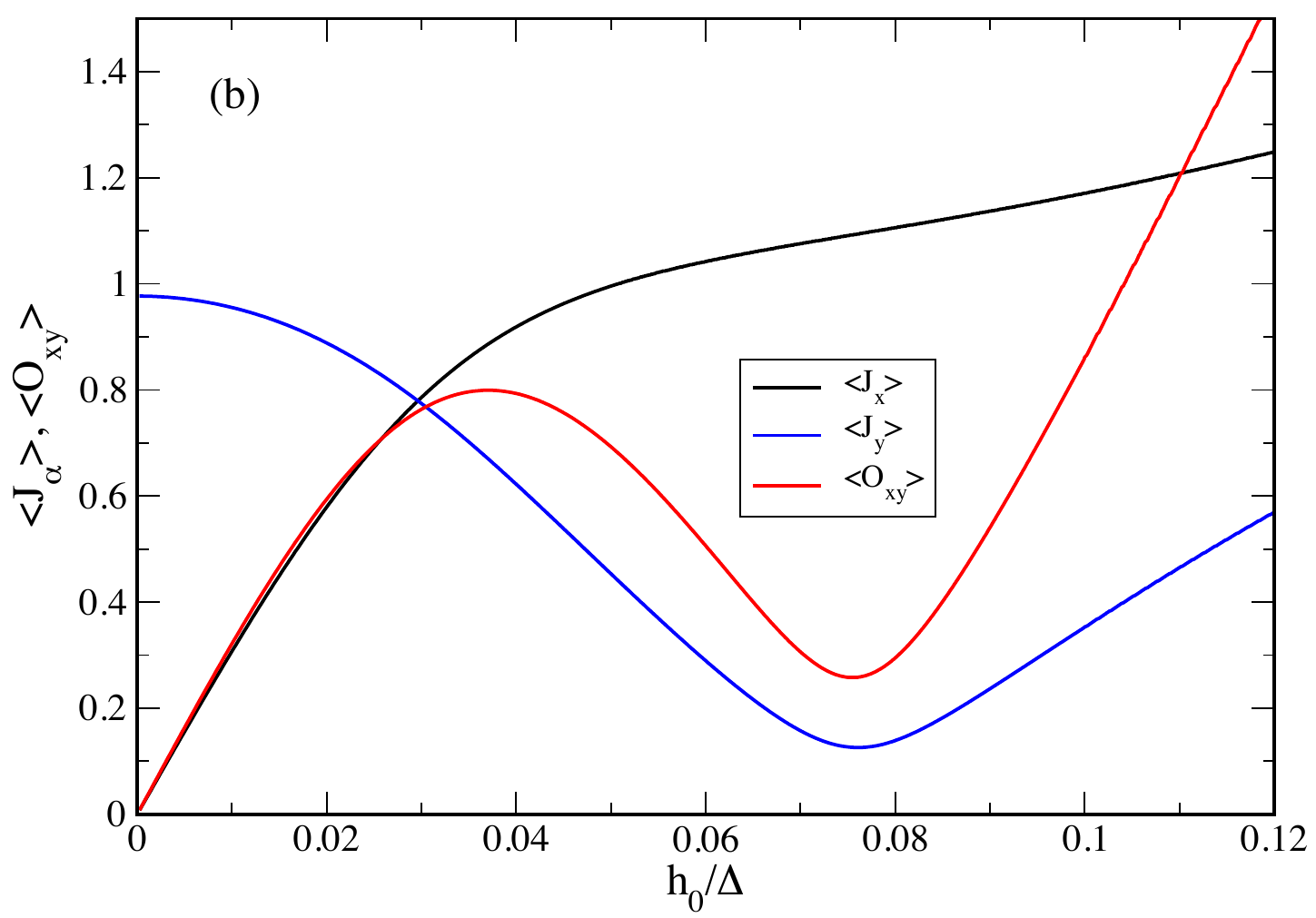}
	\caption{Dependence of homogeneous polarization $\la J_x\ra$ and order parameters $\la J_y\ra, \la O_{xy}\ra$ on the applied field. (a) For $I_m=0.019, I_Q=0.01157$ in the region with separated AFM/FIQ phases and intervening para-phase. On the upper critical field $h^+_{0\text{cr}}$ the order parameters show a first-order type jump to finite value. (b) For $I_m=0.019, I_Q=0.0117$ within merged coexistence phase. In both cases $T/\De=0.005$. For intermediate $I_Q$ corresponding to QCP endpoint see Fig.~\ref{fig:OP-crit}. }
	\label{fig:OP-h0}
\end{figure}
Finally we proceed to discuss to the ordered regimes below the critical field curves and investigate how the order parameters evolve with the field and whether it agrees with previous conjectures made from the instabilities approached from the disordered regime. The order parameters together with the homogeneous polarization $\la J_z\ra$ as obtained from the selfconsistent solution of Eq.~(\ref{eq:MFOP}) are shown in Fig~\ref{fig:OP-h0} for the two regions. (a) with separated critical fields $h^\pm_\text{cr}$ for AFM and FIQ QCP's and (b) for the coexistence case with nonzero order parameters for the whole field range. In the left part of (a) the primary AFM order parameter first induces the quadruple (red) and then at the critical AFM field $h^-_{0\text{cr}}$ forces it to drop to zero again. Then a disordered sector prevails up to $h^+_{0\text{cr}}$ where now a primary quadrupolar (red) order parameter reappears due to the large field- induced quadrupole matrix element of the ground state doublet. It again induces a secondary AFM order parameter. Therefore when progressing from $h^-_{0\text{cr}}$ to $h^+_{0\text{cr}}$ the magnetic and quadrupolar order parameter interchange their roles.

\begin{figure}
	\includegraphics[width=0.80\columnwidth]{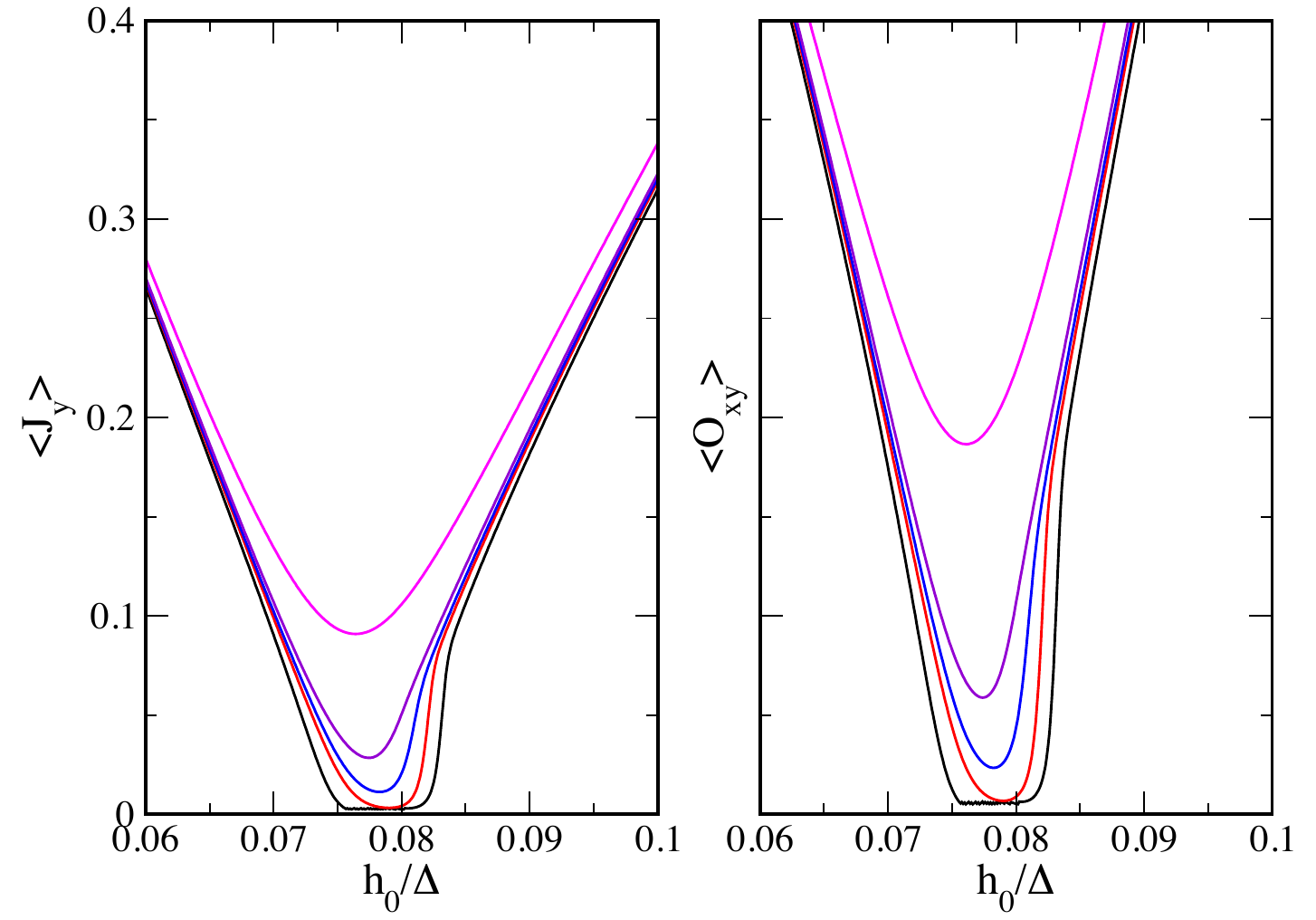}
	\caption{Enlarged dependence of homogeneous polarization $\la J_x\ra$ and order parameters $\la J_y\ra, \la O_{xy}\ra$ on the applied field in the critical QCP endpoint region of Fig.~\ref{fig:tm-h0}(b). Here $I_m=0.019$ and proceeding from lower to upper curves: $I_Q=0.01165-0.01168$. Both curves are asymmetric in $h_0$ exhibiting a first-order type jump-like behavior at the upper critical field $h_{0\text{cr}}^+$.}
	\label{fig:OP-crit}
\end{figure}
In the coexistence case (b) the critical fields cease to exist and both order parameters are finite in the whole field range. However, their pronounced dip on the previous critical field positions is still prominent and should lead to rather similar thermodynamic anomalies when crossing the dip region as compared to the case (a) when the critical fields are still present. The transition region between (a) and (b) where the upper and lower QCP's merge into an endpoint is shown in Fig.~\ref{fig:OP-crit} in an enlarged scale. The critical value for $I_Q$ where the two critical field curves touch and merge is very close to the paramagnetic calculations in Figs.~\ref{fig:tm-h0} and~\ref{fig:tm-full} that corresponds to the situation in CeRh$_2$As$_2$.

\section{Numerical treatment of C\lowercase{e}R\lowercase{h}$_2$A\lowercase{s}$_2$ with full CEF level scheme }
\label{sec:exactRPA}

\begin{figure}
	\includegraphics[width=0.8\columnwidth]{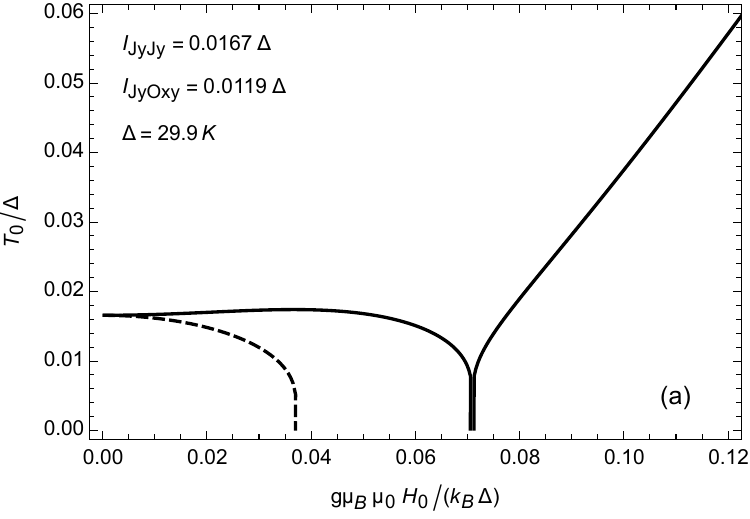}
	\includegraphics[width=0.8\columnwidth]{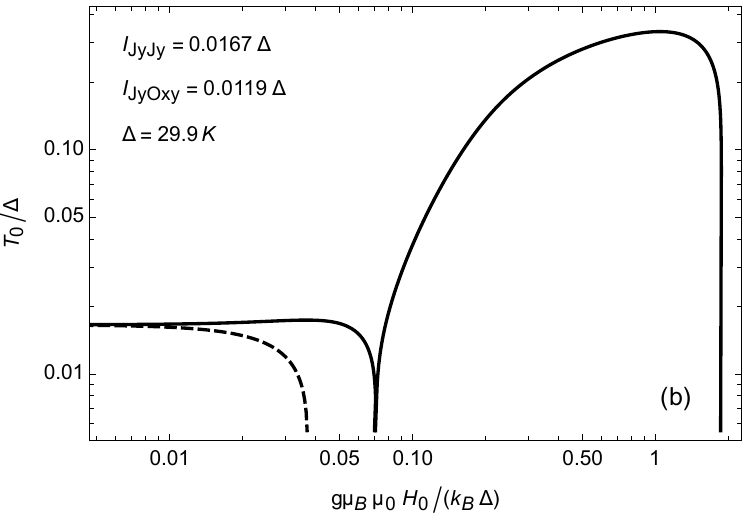}
	\caption{Phase boundary plot for \bH$_0\parallel$ a (full line) and \bH$_0\parallel$ c (broken line) as obtained from numerical calculations for the full $J=5/2$ CEF level scheme with three Kramers doublets at $0, 30, 180$ K. (a) low field regime (cf. Fig.~\ref{fig:tm-h0}(a)). (b) full field range in log-log plot. The maximum corresponds to $T_\text{cr}\simeq 0.3\De$ at $h_0/\De\simeq 1$. The CEF parameters (Sec.~\ref{sec:CEFmodel}) are taken from experiment~\cite{hafner:22}.}
	\label{fig:tm-full}
\end{figure}
The numerical evaluation of the RPA equations starting from Eqs.~(\ref{eq:baresus}) is in principle identical to the treatment of the quasi-quartet in the preceding sections, however this time involving all six crystal-field states of the $J=5/2$ multiplet. This is important as soon as temperatures and/or applied magnetic fields cannot be regarded as small compared to the CEF level splittings, in particular to $\De$.

We start with Eq.~(\ref{eq:RPAsusmat}) as before. The general form of the bare susceptibility $\chi_{AB}^0=\chi_{AB}^\text{vV}+\chi_{AB}^\text C$ we use is
\begin{equation}
\begin{split}
	\chi_{AB}^\text{vV}
	&=
	\mathop{\rm Tr}\left[M(A,B)\cdot D\right],
	\\
	\chi_{AB}^\text C
	&=
	\beta\left\{
	\mathop{\rm Tr}\left[M(A,B)\cdot P\right]
	\right.
	\\
	&\quad
	\left.
	-\mathop{\rm Tr}\left[N(A)\cdot P\right]
	\mathop{\rm Tr}\left[N(B)\cdot P\right]
	\right\}.
\end{split}
\end{equation}
For the respective components we use indices $m,n$ to label the CEF states and define $(2J+1=6)$-dimensional matrices
\begin{equation}
\begin{aligned}
	M(A,B)
	&=
	\left(m_{mn}(A,B)\right),
	\\
	m_{mn}(A,B)
	&=
	\left\langle m\left|A\right|n\right\rangle
	\left\langle n\left|B\right|m\right\rangle,
	\\
	D
	&=
	\left(d_{mn}\right),
	\\
	d_{mn}
	&=
	\frac{p_m-p_n}{E_n-E_m}\left[1-\delta(E_n-E_m)\right],
	\\
	N(A)
	&=
	\left(n_{mn}(A)\right),
	\\
	n_{mn}(A)
	&=
	\left\langle m\left|A\right|n\right\rangle,
	\\
	P
	&=
	\left(p_{mn}\right),
	\\
	p_{mn}
	&:=
	p_m\delta(E_n-E_m),
	\\
	p_m
	&:=
	\frac1Z{\rm e}^{-\beta E_m},
	\quad
	Z:=\sum_n{\rm e}^{-\beta E_n}
\end{aligned}
\end{equation}
where the expression $\langle m|A|n\rangle$ denotes a matrix element of an operator $A$ between states $|m\rangle$ and $n\rangle$ for finite molecular field $\bf h$. Here we replace the notion of ``diagonality'' with ``energy equality'' to avoid numerical issues not only at zero field (Kramers doublets) but rather also at large applied fields when the Zeeman splitting is of the order of the crystal-field splitting.

To determine the RPA phase boundary $T_\text{cr}(h)$ we use the secular equation equivalent to Eq.~(\ref{eq:singularity}). Coming from the paramagnetic side, $h$ is the total molecular field in either the $x$ direction or the $z$ direction, respectively. Therefore, as a second step, we have to determine the applied field through $h_0 = h+I_m\left\langle J_\alpha\right\rangle$ where $\alpha=x$ or~$z$ depending on the field direction, see also Eqs.~(\ref{eq:MFOPa}). Here the angular brackets denote the thermal expectation value calculated as the average over the statistical operator of the mean-field Hamiltonian. Fig.~\ref{fig:tm-full} shows the two phase boundaries for ${\bf h}_0$ parallel to $x$ (solid line) and $z$ (dashed line) determined in this way for one particular choice of parameters closely resembling 
the experimental situation (Fig.~\ref{fig:tcr-exp}) in CeRh$_\text 2$As$_\text 2$~\cite{hafner:22,semeniuk:23}.

\begin{table*}
\caption{Summary of critical and maximum quantities (units: $\De=30\;\mbox{K}$ ,T or K) obtained from
the the theoretical results in Fig.~\ref{fig:tm-full} together with
two critical fields from experiments in CeRh$_2$As$_2$~cite{hafner:22,semeniuk:23}.
The reduction factor (last column) is evidence of the Kondo screening of moments
and of similar size for both directions.
}
\vspace{0.5cm}
\centering
\begin{tabular}{c @{\hspace{4mm}} c @{\hspace{4mm}} c @{\hspace{4mm}} c}
\hline\hline
critical/extremal quantity& CEF-RPA calculation & experiment & ratio\\\hline
$h^c_{0\text{cr}}$ & $0.0375\equiv 1.95\;\mbox{T}$ &$6-7\;\mbox{T}$& $0.33$ \\
$h^a_{0\text{cr}}$ & $0.071\equiv 3.69\;\mbox{T}$ & $9\;\mbox{T}$ &$0.41$ \\
$h^a_{0\text{max}}$ & $1\equiv 52.11\;\mbox{T}$ & $-$ & $-$ \\
$T^a_\text{max}$ & $0.34 \equiv 10.2\;\mbox{K}$ & $-$ & $-$ \\
$h^a_{0\text{cr}'}$  & $1.95\equiv 96.40 ;\mbox{T}$ & $-$ & $-$ \\
\hline\hline
\end{tabular}
\label{tbl:critval}
\end{table*}
We now turn to the question to which extent the low-field approximation for the quasi-quartet model in Fig.~\ref{fig:tm-h0}(a) relates with the all-numerical calculation for the full level scheme including three Kramers doublets. This calculation is valid for any field strength and it is interesting to follow $T_\text{cr}$ to find out its possible maximum $T^a_\text{max}$ and corresponding maximum field $h^a_{0\text{max}}$ as well as the upper critical field $h^a_{0\text{cr}'}$ which should occur when the Zeeman splitting becomes comparable to $\De$ and homogeneous field polarization overwhelms the staggered order. First we show again the low field regime in Fig.~\ref{fig:tm-full}(a). It agrees well qualitatively with the quasi-quartet results in Fig.~\ref{fig:tm-h0}(a) with only minor numerical differences of critical field and quadrupolar interaction parameters. The extension to high fields is shown in a double logarithmic plot in Fig.~\ref{fig:tm-full}(b). It demonstrates that the maximum occurs at $T_\text{cr}(h^a_{0\text{max}})=T^a_\text{max}\simeq 0.34\De$ for $h^a_{0\text{max}}\simeq \De$ which is in the expected range. The very large increase of $T_\text{cr}$ of the induced quadrupolar phase compared to the zero field value of the magnetic phase is due to the large ratio of matrix elements $m'_Q/m_{a1}$.

At this point it is appropriate to estimate the absolute magnitude of critical fields, temperatures and magnetic moment in view of partly known experimental quantities. We have shown already that the anisotropy and other basic features of the calculated magnetic and quadrupolar phase diagram reproduces the empirical findings in Refs.~\onlinecite{hafner:22,semeniuk:23} and others. It must be said from the outset, however, that one cannot expect a quantitative agreement since we used a purely localized 4f electron approach. In reality the Kondo screening will have a large influence on the magnetic properties. This can be directly seen from the ordered moment $\mu=g_J\mu_\text B\la J_\al\ra_0$ ($\al=x,z$, $g_J=6/7$): with the saturation order parameter (Eq.~(\ref{eq:polpara})) $\la J_\al\ra_0=0.97$ we have $\mu=0.83\mu_B$. There are no experiments yet that have identified an ordering of the moments found in $\mu SR$ investigations~\cite{khim:24} but one should expect a strong reduction of its magnitude due to Kondo screening known also from other magnetically ordered heavy fermion compounds.

This may be concluded from the fact that the Kondo scale $T^*$ has been reported~\cite{hafner:22,christovam:24} to be of the same order of magnitude as the CEF splitting $\De$. Using the scaling factor $(g_J\mu_B)/\De=0.0192T^{-1}$ we obtain critical/maximum field and temperature values that are compared to the know experimental values in Table~\ref{tbl:critval}. It shows that the theoretical critical field values are lower than the experimental ones which may again be attributed to the large unscreened moments in the localized picture resulting in a too large Zeeman effect.

Table~\ref{tbl:critval} shows that the maximum quadrupolar ordering temperature reached at $H^a_{0\text{max}}=52\,\mbox{T}$ is $T_\text{cr}^\text{max}= 10.2\,\mbox{K}$ and the upper critical field $H^a_{0\text{cr}'}=96\,\mbox{T}$ where induced quadrupole order is finally destroyed. These exceptionally enhanced values have not yet been identified in CeRh$_2$As$_2$. We note, however, that similar large values are known from a related compound with quadrupolar order, the cubic, genuine $\Gamma_8$ quartet system CeB$_6$~\cite{shiina:97,shiina:02,thalmeier:21} which shows quadrupolar order already in zero field (and induced octupolar order at finite field) due to the absence of splitting in cubic symmetry. It has maximum values observed in pulsed-field experiments at $(40\,\mbox{T},10\,\mbox{K})$ and a upper critical field estimated to be $80\,\mbox{T}$~\cite{goodrich:04,shiina:02}. Thus the theoretical values obtained here in the localized 4f-electron approach may well give a realistic estimate in particular since the influence of Kondo screening is strongly reduced at such high fields.

\section{Summary and Conclusion}
\label{sec:summary}

In this work we have investigated the possible origin of the extremely anisotropic normal state phase diagram of tetragonal CeRh$_2$As$_2$. We use a fully localized CEF- split 4f-electron model for the $J=5/2$ multiplet of $Ce^{3+}$. We do not include the effect of Kondo screening leading to local moment reduction and heavy conduction band formation. In fact recent ARPES experiments have suggested that CeRh$_2$As$_2$ is close to the localized $4f$ limit~\cite{chen:24}. Furthermore similar examples of Ce-hexaboride and \mbox{-skutterudite} compounds have demonstrated that neglecting the Kondo screening may be an acceptable starting point for understanding major qualitative features of the $H$-$T$ phase diagram although it may be too simple to explain the quantitive aspects.

The starting point is the conjecture that the anisotropy of this phase diagram is caused by a multipolar competition of low-field magnetic dipolar and high-field induced electric quadrupolar order parameters. This competition has been investigated analytically within a simplified quasi-quartet model valid for low fields and temperatures and numerically for the complete level scheme in the full range. We employ the RPA response function technique from the disordered side and the coupled MFA for polarization and order parameters, using effective operator technique in the ordered regime. The results of analytical and numerical approach agree in the low field regime. In the high field case the latter predicts the phase boundary in a region not yet tested experimentally.

The normal state $H$-$T$ phase diagram for $\bH_0 \parallel c$ has the appearance of an antiferromagnet (as suggested by $\mu SR$ experiments~\cite{khim:24}) while for $\bH_0\parallel a$ another high field phase appears immediately after the low field phase region. Its critical temperature rises without limitation in the field range so far probed. We have shown that such behavior can be explained by the presence of easy-plane antiferromagnetic order $\la J _y\ra$ (moments perpendicular to $\bH_0\parallel a$) of $\Gamma_5$ symmetry and a field-induced quadrupolar order parameter $\la O_{xy}\ra $ belonging to $\Gamma_4$ type irreducible representation of C$_{4\text v}$. The FIQ phase appears because the mixing of $\Ga_6$ excited state into the ground state $\Ga^{(1)}_7$ where a strong matrix element $(m'_Q)$ creates a corresponding field induced quadrupolar matrix element in the ground state (absent for zero field) which increases rapidly with applied field strength $H_0$. In the paramagnetic phase this means that a mixed dipolar-quadrupolar susceptibility appears such that increasing $H_0$ leads to a divergence for the quadrupolar RPA susceptibility at the induced ordering temperature. Likewise the coupled MF equations of homogeneous polarization $\la J_x\ra$ and order parameters $\la J _y\ra$ and $\la O_{xy}\ra $ in the ordered regime show that in the low field case the primary magnetic order induces the quadrupole and vice versa in the high field phase. The two QCP's where the respective order parameters vanish enclose a disordered regime. The full level scheme calculation shows that the FIQ phase will extend to high temperatures and fields similar as observed in the true $\Gamma_8$ quartet system CeB$_6$ although in this compound it is the coexistence of primary quadrupolar and field-induced octupolar order that drives the strong increase.

The dipolar exchange and quadrupolar effective interaction determine the ordering temperatures and the critical fields. When the latter increases the disordered regime shrinks and vanishes at a quantum critical endpoint. This situation corresponds closely to the one observed in CeRh$_2$As$_2$. The interaction strengths may be characterized by dimensionless control parameters which are rather asymmetric (small for dipolar exchange and sizable but subcritical $(< 1)$ for quadrupolar interaction). This points to the fact that the zero-field AF order is driven by the $\Gamma_7^{(1)}$ ground state moments and FIQ order by the field induced moments due to $\Gamma_7^{(1)}-\Gamma_6$ mixing. If the quadrupolar control parameter would be above critical $(>1)$ the quadrupolar order would likewise appear as self-induced order already at zero field as observed e.g. in the $J=\frac{7}{2}$ compound YbRu$_2$Ge$_2$. The near equality of the two critical fields, i.e., the stability of the quantum critical endpoint may be obtained along a line in a sizable region of the control parameter plane.

\begin{table}
\caption{Matrix elements of dipolar and quadrupolar operators, see Eqs.~(\ref{eq:matdip}) to~(\ref{eq:matqua}), in the basis of the crystal-field doublets. The primes indicate matrix elements between doublets of different symmetry, $a$ and $c$ label respective in-plane and out-of-plane matrix elements, $Q$ denotes the quadrupole matrix elements. The CEF mixing angle for CeRh$_2$As$_2$ is $\theta=0.346\pi$~\cite{hafner:22}, the variation with $\theta$ is shown in Fig.~\ref{fig:matrix}.}\vspace{0.3cm}
\centering
\begin{tabular}{ccc}
\hline\hline
matrix element &CEF expression& CeRh$_2$As$_2$\\\hline
$m_{c1}$ & $1/2+2\cos2\theta$ & -0.63\\
$m_{c2}$ & $1/2$ & $0.5$ \\
$m_{a1}$ & $-(\sqrt{5}/2)\sin2\theta$ & $-0.92$ \\
$m_{a2}$ & $3/2$ & 1.5 \\
$m'_a$   & $-\sqrt{2}\sin\theta$ & $-1.25$ \\
\hline
$m_{c3}$ & $2\sin2\theta$ & $1.65$\\
$m_{a3}$ & $(\sqrt5/2)\cos2\theta$ & $-0.63$\\
$m_{a2}'$ & $\sqrt2\cos\theta$ & $0.66$\\
\hline
$m'_{Q}$ & $\sqrt{2}(\sqrt{5}\cos\theta+3\sin\theta)$ & $5.23$ \\
$m'_{Q2}$ & $-\sqrt{2}(\sqrt{5}\sin\theta-3\cos\theta)$ & $-0.83$ \\
\hline\hline
\end{tabular}
\label{tbl:matel}
\end{table}
The experimental verification of the proposed scenario of order parameters requires a diagnosis of the ordered phases by various means like neutron diffraction in external field, NMR experiments as well as resonant x-ray scattering. To map out the phase boundary in the high field regime with increased $T_\text{cr}$ ultrasonic and resistivity measurements in pulsed fields may be suitable. Based on the theoretical foundations laid in this work it will also be possible to calculate the field- and temperature dependence of some of the experimental quantities, in particular the evolution of their anomalies along the phase boundaries. These quantities are specific heat, symmetry elastic constants, magnetostriction and thermal expansion. Furthermore dynamical properties like collective modes in the ordered regime, in particular for high fields can be calculated within the theory developed here. These further investigations will be reported elsewhere.

\appendix
\section{The multipolar operators in the $J=\frac{5}{2}$ CEF state basis of CeRh$_2$As$_2$}
\label{sec:appCEF}

\begin{figure}
\includegraphics[width=\columnwidth]{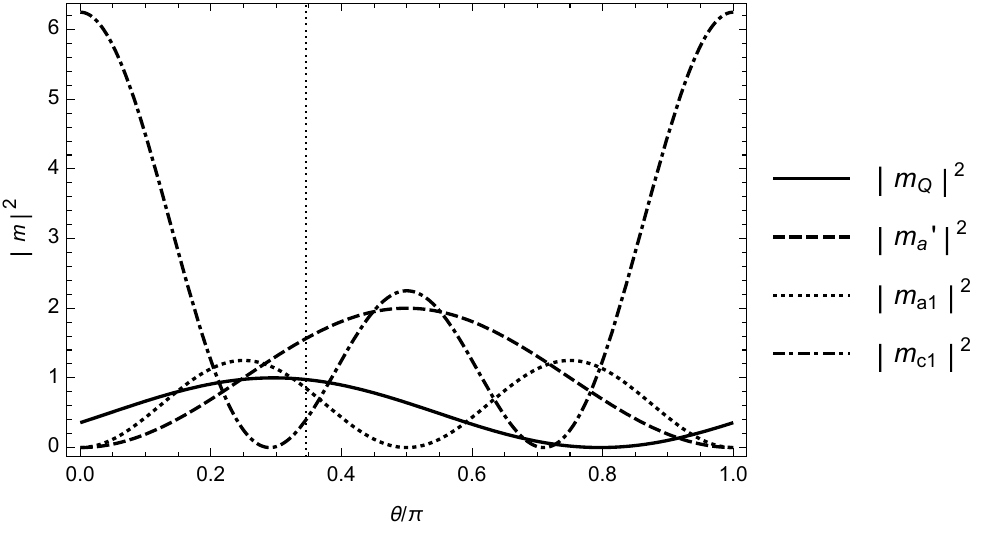}
\caption{Selected nonzero matrix elements between the components $|\Gamma_7^{(1)}\rangle$ and $\Gamma_6\rangle$ of the quasi-quartet for the quadrupole component $O_{xy}$ and the dipole operators as a function of the mixing angle $\theta$. The quadrupole matrix element (solid line) is normalized to its maximum value $|m_Q(\theta_\text{max})|^2=28$ with $\theta_\text{max}=2\tan^{-1}\left[(1/3)\sqrt{19-2\sqrt{70}}\right]\approx0.296\pi$. The thin vertical line denotes $\theta=0.346\pi$ taken from experiment~\cite{hafner:22}.}
\label{fig:matrix}
\end{figure}
Here we give details about the multipolar operators and their matrix elements in the representation using the CEF states of Eq.~(\ref{eq:CEFstate}) as basis. Table~\ref{tbl:matel} holds a compilation of the general expressions (depending on CEF mixing angle $\theta$) of the nonzero matrix elements of dipolar and quadrupolar operators, for the latter see Table~\ref{tbl:quadop}. It is obvious that the quadrupolar matrix element $m'_Q$ for CeRh$_2$As$_2$ is comparatively strong enabling the pronounced field induction of the $O_{xy}$ quadrupole. The $\theta$ variation of (non-constant) matrix elements is shown in Fig.~\ref{fig:matrix}. The relevant multipole order parameters necessary for the analysis are represented by $6\times6$ matrices using the row and column sequences $\left\{|\Gamma_{7+}^{(1)}\rangle,|\Gamma_{7-}^{(1)}\rangle,|\Gamma_{6+}\rangle,|\Gamma_{6-}\rangle,|\Gamma_{7+}^{(2)}\rangle,|\Gamma_{7-}^{(2)}\rangle\right\}$:
\begin{equation}
\begin{gathered}
	\mbox{dipolar $\Gamma_5$ operators:}
	\\
	J_x
	=
	\left(
	\begin{array}{cccc|cc}
		0 & m_{a1} & 0 & m_a' & 0 & m_{a3} \\
		m_{a1} & 0 & m_a' & 0 & m_{a3} & 0 \\
		0 & m_a' & 0 & m_{a2} & 0 & m_{a2}' \\
		m_a' & 0 & m_{a2} & 0 & m_{a2}' & 0 \\
		\hline
		0 & m_{a3} & 0 & m_{a2}' & 0 & -m_{a1} \\
		m_{a3} & 0 & m_{a2}' & 0 & -m_{a1} & 0 \\
	\end{array}
	\right)
	\\[\baselineskip]
	J_y
	=
	{\rm i}
	\left(
	\begin{array}{cccc|cc}
		0 & -m_{a1} & 0 & m_a' & 0 & -m_{a3} \\
		m_{a1} & 0 & -m_a' & 0 & m_{a3} & 0 \\
		0 & m_a' & 0 & -m_{a2} & 0 & m_{a2}' \\
		-m_a' & 0 & m_{a2} & 0 & -m_{a2}' & 0 \\
		\hline
		0 & -m_{a3} & 0 & m_{a2}' & 0 & m_{a1} \\
		m_{a3} & 0 & -m_{a2}' & 0 & -m_{a1} & 0 \\
	\end{array}
	\right)
\end{gathered}
\label{eq:matdip}
\end{equation}
\begin{equation}
\begin{gathered}
	\mbox{dipolar $\Gamma_2$ operator:}
	\\
	J_z
	=
	\left(
	\begin{array}{cccc|cc}
		m_{c1} & 0 & 0 & 0 & m_{c3} & 0 \\
		0 & -m_{c1} & 0 & 0 & 0 & -m_{c3} \\
		0 & 0 & m_{c2} & 0 & 0 & 0 \\
		0 & 0 & 0 & -m_{c2} & 0 & 0 \\
		\hline
		m_{c3} & 0 & 0 & 0 & 1-m_{c1} & 0 \\
		0 & -m_{c3} & 0 & 0 & 0 & -(1-m_{c1}) \\
	\end{array}
	\right)
\end{gathered}
\end{equation}
\begin{equation}
\begin{gathered}
	\mbox{quadrupolar $\Gamma_4$ operator:}
	\\
	O_{xy}
	=
	{\rm i}
	\left(
	\begin{array}{cccc|cc}
		0 & 0 & -m_Q' & 0 & 0 & 0 \\
		0 & 0 & 0 & m_Q' & 0 & 0 \\
		m_Q' & 0 & 0 & 0 & -m_{Q2}' & 0 \\
		0 & -m_Q' & 0 & 0 & 0 & m_{Q2}' \\
		\hline
		0 & 0 & m_{Q2}' & 0 & 0 & 0 \\
		0 & 0 & 0 & -m_{Q2}' & 0 & 0 \\
	\end{array}
	\right)
\end{gathered}
\label{eq:matqua}
\end{equation}
Within the quasi-quartet model subspace the respective top left $4\times4$ blocks (separated by thin lines) represent the multipoles within the corresponding sequence $\left\{|\Gamma_{7+}^{(1)}\rangle,|\Gamma_{7-}^{(1)}\rangle,|\Gamma_{6+}\rangle,|\Gamma_{6-}\rangle\right\}$ of quasi-quartet states. Since we want to calculate the $H$-$T$ phase diagram we also need these operators expressed in terms of the eigenstates in an external field corresponding to the total local Hamiltonian for each site. For $\bH\parallel c$ there is no change while for $\bH\parallel a$ the multipoles in quasi-quartet subspace are given in Eqs.~(\ref{eq:matfield-Jx}) and~(\ref{eq:matfield-Jy}) and~(\ref{eq:matfield-O}).

\section{Expressions for effective operators in the effective ground state doublet}
\label{sec:appeff}

For the calculation of order parameters in Sec.~\ref{sec:OP} we need to compute the diagonal elements of multipole operators with the eigenstates of the effective Hamiltonian Eq.~(\ref{eq:HAMeff}) described by Eq.~(\ref{eq:Umat}) which contain the admixture effects with the upper $\Gamma_6$ levels. For an operator $A=J_x,J_y,O_{xy}$ this is achieved by defining the dressed operator $\cal A\rm_\text{eff}$ ($H_1$ is defined in Eq.~(\ref{eq:HMF})):
\begin{equation}
\begin{aligned}
	&
	\la k | \cal A\it_\text{eff}|l\ra		
	=\\
	&\quad
	A_{kl}-\frac{1}{\De}\sum_\mu
	\left[\la k|H_1|\mu\ra\la\mu|A | l \ra +
	\la k| A|\mu\ra\la\mu| H_1| l \ra\right]
\end{aligned}
\end{equation}
where $k,l$ runs over the unperturbed $|1\pm\ra$ ground state doublet and $\mu$ over the unperturbed $|2\pm\ra$ excited doublet states of the quasi quartet. The diagonal matrix elements in the split ground state doublet of the MF effective Hamiltonian (Eq.~(\ref{eq:HAMeff})) are then given by 
\begin{equation}
\hspace{-0.5cm}
\la \psi_{n\lam}|A|\psi_{n\lam}\ra=\sum_{kl}U^{\lam*}_{nk}U_{nl}^{\lam}\la k|\cal A\it_\text{eff}|l\ra =tr(W^{\lam}_n\cal A\it_\text{eff})
\end{equation}
where the $|\psi_{n\lam}\ra$ are the column vectors in Eq.~(\ref{eq:Umat}). Furthermore we defined the Hermitian matrix $\{W^{\lam*}_n\}_{kl}=U^{\lam*}_{nk}U^\lam_{nl}=\{W^{\lam}_n\}_{lk}$ or explicitly $(n=\pm)$
\begin{equation}
W^\lam_n=\fs
 \begin{pmatrix}
1& -ne^{i\lam\phi} \\
-ne^{-i\lam\phi}&1\\
\end{pmatrix}
\end{equation}
Using these diagonal elements and the thermal occupations of split ground state levels (Eq.~(\ref{eq:spliteff})) the selfconsistent order parameter equations (Eq.~(\ref{eq:MFOP})) may be derived.

\begin{acknowledgments}
The authors thank Manuel Brando and Seunghyun Khim for helpful discussions and sharing their experimental results, in particular those adapted from Refs.~\cite{hafner:22,semeniuk:23} and presented in Fig.~\ref{fig:tcr-exp}.
\end{acknowledgments}

\bibliography{cerh2as2}
\end{document}